\documentclass[prd,aps,preprint]{revtex4}
\usepackage{graphicx}
\begin{document}
\tolerance=5000
\def\be{\begin{equation}}
\def\ee{\end{equation}}
\def\bea{\begin{eqnarray}}
\def\eea{\end{eqnarray}}
\def\ii{\'{\i}}
\def\bi{\bigskip}
\def\be{\begin{equation}}
\def\en{\end{equation}}
\def\bq{\begin{eqnarray}}
\def\eq{\end{eqnarray}}
\def\noi{\noindent}

\title{Time and ``angular" dependent backgrounds from stationary 
axisymmetric solutions}
\author{Octavio Obreg\'on}
\email{octavio@ifug3.ugto.mx}
\affiliation
   {Instituto de F\'\i sica de la Universidad de Guanajuato \\
                    P.O. Box E-143, 37150 Le\'on Gto., M\'exico}
\author{Hernando Quevedo}
\email{quevedo@physics.ucdavis.edu}
\affiliation{Instituto de Ciencias Nucleares\\
Universidad Nacional Aut\'onoma de M\'exico\\
A.P. 70-543,  M\'exico D.F. 04510, M\'exico\\}
\affiliation{Department of Physics \\
University of California \\
Davis, CA 95616}
\author{Michael P. Ryan}
\email{ryan@nuclecu.unam.mx}
\affiliation{Instituto de Ciencias Nucleares\\
Universidad Nacional Aut\'onoma de M\'exico\\
A.P. 70-543,  M\'exico D.F. 04510, M\'exico\\}
\date{\today}

\begin{abstract}
\setlength{\baselineskip}{.5cm}
Backgrounds depending on time and on ``angular" variable, namely 
polarized and unpolarized
$S^1 \times S^2$ Gowdy models, are generated as the sector inside the 
horizons of the manifold corresponding to axisymmetric solutions.
As is known, an analytical continuation of ordinary $D$-branes, $iD$-branes
allows one to find $S$-brane solutions.  Simple models have been constructed 
by means of analytic continuation of the Schwarzchild and the Kerr 
metrics. The possibility of studying the
$i$-Gowdy models obtained here is outlined with an eye toward seeing if
they could represent some kind of
generalized $S$-branes depending not only on time but also on an ``angular''
variable.
\end{abstract}

\maketitle
\setlength{\baselineskip}{1\baselineskip}

\newpage

\section{INTRODUCTION}

For some time it has been known that the 
static solution with spherical symmetry, the Schwarzschild black hole solution,
\begin{equation} 
ds^2 =-(1-\frac{2M}{r})dt^2+\frac{1}{1-\frac{2M}{r}}dr^2 + r^2 (\sin^2 \theta
d\varphi^2 + d\theta^2)
\end{equation}
in the region $r>2M$, becomes a cosmological model as one  
crosses the horizon. What was the ``radial" direction becomes timelike,
and the timelike direction becomes spacelike. In this case the 
corresponding cosmological model is the well known Kantowski-Sachs model 
\cite{kansach}.
\begin{equation} 
ds^2 =
-\frac{1}{\frac{2M}{t}-1}dt^2 
+ (\frac{2M}{t}-1)dr^2
+ t^2 (\sin^2 \theta
d\varphi^2 + d\theta^2),
\end{equation} 

The singularity at $t=0$ is the curvature singularity of Schwarzschild $(r=0)$,
where the curvature is infinite, but the singularity at $t=2M$ is just a    
lightlike surface where the curvature is regular and we pass from the  
Kantowski-Sachs region to the Schwarzschild region. We need two copies of 
each of these regions to describe the complete casual structure of this spacetime.

In string theory one of the important open problems is the correct treatment
of time-dependent backgrounds. In \cite{sbranes} $S$-branes were first introduced.
They are objects arising when Dirichlet boundary conditions on open strings
are imposed on the time direction.  An $S$-brane is a topological defect
\cite{topdef}, all of whose longitudinal dimensions are spacelike, and
consequently they exist only for a moment of time.  $S$-branes have been
found as explicit time-dependent solutions of Einstein's equations (dilaton
antisymmetric tensor fields are also included in some models) \cite{timedep}
in the same way as black-hole-like 
solutions correspond to $p$-brane solutions. 
Some of these solutions are actually best thought of as an analytical 
continuation of ordinary D-branes, or $i$ D-branes, for short \cite{pbranes}.

A simple model has been considered in \cite{ibranes},
using only the 4D Einstein equations 
in vacuum. 
The Schwarzschild-Kantowski-Sachs model has been utilized to nicely define a  
simple $4D$ model for an $S$-brane. By analytic continuation the spherical space
$(k = 1)$ is transformed into a hyperbolic space $(k = -1)$ to obtain the
hyperbolic symmetry
$SO(2,1)$, as suggested by the fact that S-branes are kinks in time. Beginning with
the Schwarzschild solution, these authors performed the transformation 
$t\to ir$, $r\to it$, $\theta \to i\theta$, $\varphi\to i\varphi$ and   
$M\to iP$. By these means they were able to obtain a rotated Penrose diagram
for the $k=0, -1$ S-brane solution with well-defined time-dependent regions.

As mentioned above, what in Ref. \cite{samos}
 was called ``horizon methods" of generating 
cosmological solutions has a long history, though it never seems thought 
of as a ``method".  The idea of this method is to reinterpret a part of a known 
manifold as a cosmological solution of Einstein's equations.  For static and 
stationary axisymmetric solutions which have horizons,
as one crosses the horizon what
was the ``radial" direction becomes timelike, and the timelike
direction becomes spacelike, and the model becomes a time (and ``angular'')
dependent cosmological model.

In the Kantowski-Sachs model we see for the first time a problem of 
global topology.  If we insist that our model have a closed topology, 
we can achieve this by compactifying in the $r$-direction, but if one 
crosses the horizon once more, the ``Schwarzschild" portion is closed in 
the time direction, which would allow closed timelike lines.  This problem 
has been noted in several ``black-hole" -cosmology pairs, the 
most notable example being the Taub-NUT manifold \cite{taubnut}.  
Various metrics, 
including that mentioned above, as well as topological problems are discussed 
in more detail in Ref. \cite{samos}.

There has been  an enormous amount of work on exact solutions of 
``black-hole" type axisymmetric metrics, and a large number of axisymmetric
solutions with horizons have been found.  It is natural to ask what kind of 
time and ``angular" dependent backgrounds are generated as the portion of 
these manifolds inside their horizons.  Since axisymmetric metrics are 
characterized by two commuting Killing vector, one timelike and the other, 
associated with a symmetry about an axis, spacelike, it is not 
surprising that the corresponding models are the two-Killing-vector 
models studied exhaustively by Gowdy \cite{5,6}.  Since these ``black-hole" 
models are usually assumed to have compact surfaces that have an $S^2$ topology,
and one compactifies the new $t = $ constant surfaces by identification, they 
will be Gowdy models with an $S^1 \times S^2$ topology.

The paradigm for such models is the Kerr metric, whose associated 
cosmological  model was presented in Ref. 
\cite{prd}.  There are several features of this model 
that are common to many of the axisymmetric solutions.  One is that there
are two horizons, an outer horizon and an inner one.    The time and 
``angular" dependent Gowdy model is represented by the region between these two 
horizons.  Inside the inner horizon the light cones have changed in such 
a way that we have a ``black-hole" type of solution.  A second feature is that 
there are no curvature singularities in the region between these two 
horizons, which would have been an interesting example of an inhomogeneous 
singularity.  A third feature is that the apparent singularities of the Gowdy
model are only horizons. This feature is not general, as we will see below.

The no-hair theorems show that only the Kerr family of metrics can represent
true black holes.  The other members of the enormous zoo of black-hole-like
solutions are untenable as black hole models because they contain curvature 
singularities on or outside of the outer horizon.  This fact does not affect
the Gowdy interiors, but instead makes them more interesting models.  Another 
point about the ``black hole" solutions is that many of them have been found 
as solutions of the Ernst equation.  The simple fact that we have passed 
inside the horizon does not make the Ernst equation invalid, and one might 
expect that cosmological models with two commuting Killing vectors could 
be generated using the same techniques that were useful in the stationary 
axisymmetric case.
This possibility will be discussed in detail in a 
forthcoming paper by the same authors. Recently, it was shown that
particular Gowdy models can be generated from the data given on a specific
hypersurface by applying solution generating techniques \cite{aah}.
In the present article we will
only make mention of some points related to this concept.  As we mentioned
above, the entire zoo of axisymmetric solutions with horizons could be 
laboriously converted to Gowdy models and their features studied.  In 
this paper we only plan to give a few of the more interesting solutions.
These solutions can be broken down into three categories, each one with a 
representative metric or class of metrics.  These three are:

\bi

\begin{center}
  1) Simple solutions --- the Zipoy-Voorhees metric,\\
  2) More complicated Kerr-like solutions --- Tomimatsu-Sato metrics,\\
  3) Complicated curvature-singularity-horizon behavior on the ``horizons"\\
  --- the Erez-Rosen metric.\\
\end{center}

\bi

  The models we will present correspond to each of these three cases.
They are time and ``angular" dependent backgrounds, Gowdy models,
which we will describe in this article. It is not the purpose of this work
to analyze the possibility of defining the corresponding models as 
$i$-models as has been done with $iD$-branes in \cite{pbranes} and for a
simple $4D$ model in \cite{ibranes}, and recently for the Kerr metric 
in \cite{tasi} and \cite{wang}.  An analysis of generalized S-brane 
solutions corresponding to the three categories of metrics considered 
here is in progress and will be reported on elsewhere.
 In general we expect these $S$-brane backgrounds 
to be gravitational fields
rather than simple counterparts of homogeneous cosmologies. Gowdy models are
the simplest example of a true field theory.
This has always been one of the major uses of Gowdy models, that is, as a
gravitational field theory that (in the polarized case) has simple exact
solutions.

In Section II we present some general considerations for the 
transition from axisymmetric static, stationary solutions to $S^1 \times 
S^2$ time and ``angular" dependent Gowdy models in the context of the 
horizon method.  In sections III, IV and V we will present the models
corresponding to each of these metrics mentioned above, and will analyze 
the behavior of the relevant metric functions, especially near the singularities 
and horizons.  Section VI is devoted to conclusions and suggestions 
for further research.

\section{GENERAL CONSIDERATIONS}

Axisymmetric solutions are often given in Lewis-Papapetrou
form. Here we will use prolate spheroidal coordinates  
where the metric takes the form
\bea
ds^2 = & - &f (dt- \omega d\varphi)^2 
+ f^{-1} (x^2 -1) (1-y^2) d\varphi^2\nonumber \\
       && + f^{-1} e^{2\gamma} (x^2-y^2)
        \left[ {dx^2\over x^2-1} + {dy^2\over 1-y^2}\right] , 
\label{met1}
\eea
where $f, \omega,$ and $\gamma$ are functions of $x$ and $y$. One
sometimes writes $f = A/B$, where $A$ and $B$ are also functions
of $x$ and $y$. 

The variable $x$ is the ``radial" coordinate, and
the outer and inner ``horizons" (the reason for the quotation
marks will become obvious later) are at $x = \pm 1$. For ``black
hole" solutions one takes $x > 1$. The quantity $y$ is the
``angular" variable, and it is usual to make the transformation $y
= \cos\theta$, where $\theta$ is the ordinary polar angle. The
cosmological sector of these metrics is the region where $-1 < x <
+1$. In order to make contact with previous Gowdy formulations,
we will define $x = \cos(e^{-\tau})$ in the cosmological region.
In this region the term $dx^{2}/(x^{2} - 1)$ changes sign, which
allows us to interpret $\tau$ as a time coordinate. With these
changes, the metric (\ref{met1}) between its horizons takes the form
for unpolarized $S^{1} \times S^{2}$ Gowdy models that has been
used by several authors \cite{11,10},
\bea ds^2 &=& e^{-\lambda /2} e^{\tau /2} (-e^{-2\tau} d\tau^2 + d\theta^2)
\nonumber \\
&& + \sin (e^{-\tau}) [e^P d\chi^2 + 2e^P Q d\chi d\varphi +(e^P Q^2 +
e^{-P} \sin^2 \theta ) d\varphi^2], 
\label{sls2}
\eea 
where $\chi$ is the $t$ of (\ref{met1}) and $\partial/\partial\chi$ is 
now supposed to be a spacelike direction, and we
have compactified in the $\chi$ direction by supposing that 
$0 \leq \chi \leq 2\pi$, with zero and $2\pi$ identified.
The functions in this metric can be identified with 
$A,\ B,\ \gamma$ and $\omega$ of (\ref{met1}) by
\be
e^{-\lambda /2} e^{\tau /2} = {B\over A} e^{2\gamma}(\cos^2 e^{-\tau} - \cos^2\theta)
\label{lambda}
\ee
\be
e^P = - {A\over B\sin e^{-\tau}}\ , \qquad Q = -\omega \ .
\label{pandq}
\ee
The form of the $d \varphi^{2}$ term is due to the fact that the
two-metric in $d\chi$ and $d\varphi$ must have determinant 
$\sin^2\theta \sin^{2}( e^{-\tau})$
for the metric (\ref{sls2}) to be a solution of the Einstein equations,
a condition that (\ref{sls2}) obeys. 

There are a number of points that
we will see in all of the three cases discussed below. The most
important point is the sign of $e^{P}$ in (\ref{sls2}). In order to
have the correct signature in (\ref{met1}), it is assumed that for $x >
1,\  A/B$ is positive, and for $P$ to be real in (\ref{sls2}), either
$A$ or $B$ must change sign at $x = 1$. While this is the case for
some of the metrics we will study, it is not true for all of them.
However, the equations for $P$ and $Q$,
\be P_{,\tau \tau} - e^{-2\tau} {{(\sin \theta P_{,\theta})_{,\theta}}\over
{\sin \theta}} - e^{-2\tau} - {{e^{2P}}\over {\sin^2 \theta}}[ (Q_{,\tau})^2
- e^{-2\tau} (Q_{,\theta})^2] - [e^{-\tau} \cot (e^{-\tau}) - 1]P_{,\tau} =
0, \label{eqp}\ee 
\be Q_{,\tau \tau} - e^{-2\tau} Q_{,\theta \theta} -
e^{-2\tau}\cot \theta Q_{,\theta} + 2(P_{,\tau} Q_{,\tau} -
e^{-2\tau} P_{,\theta} Q_{,\theta})- [e^{-\tau} \cot (e^{-\tau}) - 1]
Q_{,\tau} = 0. \label{eqq}\ee 
depend on $P$ only through $e^{2P}$ and derivatives of $P$ which
are invariant under the addition of $i\pi$ to $P$, so
if $A/B$ does not change sign, we can add $i\pi$ to $P$ and take
$e^{P} = \vert A/B\vert / \sin(e^{-\tau})$ (the variable $\tau$ runs
from $-\ln \pi$ to $+\infty$, so $\sin(e^{-\tau})$ is always positive).
Since the equation for $Q$ is invariant under a change of sign of
$Q$, we may take $Q = \pm \omega$ as we wish. Since the equations
for $\lambda$ (see Ref. \cite{10})
\bea &&\cot (e^{-\tau}) \lambda_{,\theta} - 2 e^{\tau}(P_{,\tau} P_{,\theta} +
e^{2P} {{Q_{,\tau} Q_{,\theta}}\over {\sin^2 \theta}}) \nonumber \\ 
&& + \cot \theta [-e^{\tau} \lambda_{,\tau} + 2e^{\tau} P_{,\tau} + e^{\tau}
+ 2\cot (e^{-\tau})] = 0, \label{lam1}\eea 
\bea && \cot (e^{-\tau}) (\lambda_{,\tau} - 1) - e^{\tau}[(P_{,\tau})^2 +
e^{-2\tau}(P_{,\theta})^2] - e^{\tau}{{e^{2P}}\over {\sin^2 \theta}}
[(Q_{,\tau})^2 + e^{-2\tau}(Q_{,\theta})^2]\nonumber \\ 
&& + e^{-\tau}[\cot^2 (e^{-\tau}) + 4] + e^{-\tau}(-\cot \theta
\lambda_{,\theta} + 2\cot \theta P_{,\theta}) = 0. \label{lam2}\eea 
only depend on $\lambda$ through its
derivatives, we may also take
\be
e^{-\lambda /2} e^{\tau /2} = \left | {B\over A} e^{2\gamma}\right |
|\cos^2(e^{-\tau}) - \cos^2\theta|.
\label{lameq}
\ee
For all of the metrics we will investigate, $e^{2\gamma}$
is proportional to $A$, and $B$ is positive definite (except at $x
= \pm 1$ and $\theta = 0, \pi,$ where the usual coordinate
singularity of polar coordinates occurs), so the absolute value of
$(B/A)e^{2\gamma}$ is not needed. 

A second problem with the $\tau$ and $\theta$ coordinates is that 
there might be at least a coordinate singularity at $\pm \cos(e^{-\tau})
= \cos\theta$. Of course, this depends on the form of $e^{2\gamma} A/B$.
In all of the cases we will consider,
$e^{2\gamma}$ is proportional to a power of 
$\cos^2(e^{-\tau}) - \cos^2\theta$, while $A/B$ times the rest of $e^{2\gamma}$
is regular and non-zero in the cosmological region.
In most cases the power of  $[\cos^2(e^{-\tau}) - \cos^2\theta]$ in
(\ref{lameq}) (Kerr and Erez-Rosen are the only exceptions) is
not equal to zero, so $e^{-\lambda/2}$ is singular on a surface in the 
cosmological region. It can be shown by 
an explicit coordinate transformation
given in the Appendix that for any power of 
$[\cos^2(e^{-\tau}) - \cos^2\theta]$,   this singularity
is only a coordinate effect. 
In the next sections we will study explicit examples for the 
application of this procedure.

\section{The Zipoy-Voorhees metric}

The Zipoy-Voorhees \cite{zv} metric in Lewis-Papapetrou form and 
prolate spheroidal coordinates has the form of (\ref{met1}) with
\be
\omega = 0 , \qquad f=\left({x-1\over x+1}\right)^\delta \ ,
\qquad e^{2\gamma}= \left({x^2-1\over x^2-y^2}\right)^{\delta^2} \ ,
\ee
where the constant parameter $\delta$ lies in the range $- \infty
< \delta < + \infty$ with no other restrictions, which implies
that we can take
\be
A = (x^2-1)^\delta \ , \qquad B=(x+1)^{2\delta} \ .
\ee
Between the inner and outer horizons we take $x = \cos(e^{-\tau})$,
and we have
\be
A = (-1)^\delta \sin^{2\delta} (e^{-\tau}) \ , \qquad 
B = (1+ \cos e^{-\tau}) ^{2\delta} \ .
\ee
Since in this ``polarized'' case ($Q = 0$) the equation for $P$ only
depends on derivatives of $P$, and we can add $-i \pi\delta$ to
$P$ and still have a solution, so we find that
\be
e^P = {\sin^{2\delta -1} (e^{-\tau}) \over (1+\cos e^{-\tau})^{2\delta} }\
\label{zv}
\ee
for any $\delta$ is a Gowdy solution, which can be seen by
substituting this expression into (\ref{sls2}) with $Q = 0$. In fact,
(\ref{zv}) is the general solution (up to a trivial multiplicative
constant) of (\ref{sls2}) for $P$ independent of  $\theta$, as can be
shown by quadratures. This solution can also be written as $P =
-2\delta Q_{0} (\cos e^{-\tau})/\sin e^{-\tau}$, where $Q_{0}$ is a
Legendre function of the second kind, a form well known in the $x
> 1$ region.
The expression for
$\lambda$ in this case is given by
\be
e^{-\lambda /2} e^{\tau /2} = {(1+\cos e^{-\tau})^{2\delta}\over 
(\sin^2 e^{-\tau})^{\delta-\delta^2} } \vert           
\cos^2(e^{-\tau}) - \cos^2\theta \vert^{1-\delta^2} \ .
\ee
The expression multiplying the power of $\mid \cos^{2}e^{-\tau} -
\cos^{2} \theta\mid$ is analytic and non-zero between the two
``horizons" and the singularity where $\mid\cos^{2} e^{-\tau} -
\cos^{2} \theta\mid = 0$ may be removed by the coordinate
transformation of the Appendix.
This solution is the
simplest example of a Gowdy metric obtainable from an axisymmetric
solution. 

The curvature singularities of this spacetime can be found by analyzing
the Kretschman scalar, $K= R_{\alpha\beta\mu\nu}R^{\alpha\beta\mu\nu}$, 
which in this case can be written as
\be
K= 16 \delta^2 \left({\cos e^{-\tau} - 1 \over \cos e^{-\tau} + 1}\right)^{2\delta}
{ (\cos^2 e^{-\tau} -\cos^2\theta)^{2\delta^2 -3}\over 
  (\cos^2 e^{-\tau} -1)^{2\delta^2+2} } L(\tau,\theta) \ , \label{zvk}
\ee
with
\be
L(\tau,\theta) = 3 (\cos e^{-\tau} -\delta)^2 (\cos^2 e^{-\tau}-\cos^2\theta)
+(\delta^2-1)\sin^2\theta\, [\delta^2 -1 + 3\cos e^{-\tau}\, (\cos e^{-\tau} - \delta)]
 \ .
\ee
In the special case $\delta=1$, in which we recover the Kantowski-Sachs model,
there exists only one singularity at $\tau= -\ln\pi$. It is interesting
to mention that for $\delta=-1$, the only singularity is situated at 
$\tau\rightarrow\infty$. Whereas in the Kantowski-Sachs model the singularity
at $\tau= -\ln\pi$ corresponds to a Big Bang at the origin from which the universe
expands forever free of singularities, in the dual case ($\delta = -1$) 
the universe possesses a regular origin and evolves asymptotically $(\tau\rightarrow
\infty)$ into a Big Crunch curvature singularity.

Consider now the case $\delta\neq 1$. From Eq. (\ref{zvk}) we can see that 
there is a singularity at $\cos e^{-\tau} = 1$ for $2\delta^2-2\delta+2>0$
and at $\cos e^{-\tau} = - 1$ for $2\delta^2+2\delta+2>0$. Consequently, 
there exist true singularities at $\cos e^{-\tau} = \pm 1$ for any real 
values of $\delta$. The apparent singularity at $\cos e^{-\tau} = \pm \cos\theta$,
$(\delta^2> 3/2)$, can be removed by means of the coordinate transformation of 
the Appendix for any values of the angle $\theta$ with $\cos\theta \neq \pm 1$. 
When $\cos\theta = \pm 1$ we return to the latter case. 

We have shown that the outer and inner ``horizons"
are actually surfaces of infinite curvature (naked singularities),
and they make this metric untenable as a black hole model because
it violates cosmic censorship, but between the ``horizons" it
represents a perfectly viable time and ``angular'' dependent 
background.

\section{The Tomimatsu-Sato metrics}

The Tomimatsu-Sato metrics \cite{ts} are an infinite family of metrics with
a parameter similar to the $\delta$ of the Zipoy-Voorhees metrics.
In this case, as we will see, these models give ``unpolarized'' ($Q \neq 0$)
Gowdy models between their horizons. For unpolarized Gowdy models Eqs.
(\ref{eqp}) and (\ref{eqq}) contain $e^{2P}$ as well as derivatives of $P$, so if we
want (as we will for some of the metrics) to change $P$ by 
adding $-i\pi\delta$ as we did for the Zipoy-Voorhees metrics, 
then there is an
additional restriction from the fact that the equations are only
invariant if $2i\pi  \delta$ is an integer multiple of $2 i\pi$,
that is, $\delta$ an integer. The Tomimatsu-Sato solutions
all have the equivalent of $\delta$ an integer, so we can always use them
to create backgrounds depending on an ``angular'' variable and time.

In the Lewis-Papapetrou form, all these metrics have
\begin{equation}
A = A(x,y), \qquad B = B(x,y),
\end{equation}
where $A$ and $B$ are polynomials in $x$ and $y$.  We also have
\begin{equation}
\omega = 2\delta\frac{q}{p} (1 - y^2) \frac{C}{A},
\end{equation}
where $C$ is a polynomial in $x$ and $y$, and $p$ and $q$ are numerical
constants that obey $p^2 + q^2 = 1$. Finally, we have
\begin{equation}
e^{2\gamma} = \frac{1}{p^{2\delta} (x^2- y^2)^{\delta^2}} A.
\end{equation}

Yamazaki and Hori \cite{yam} have given expressions for $A, B, C$ for the
entire infinite family.  In the original article of Tomimatsu and Sato
$A,B$ and $C$ were
given for $\delta=1,2,3,4$.  Since the maximum powers of the polynomials 
$A$ and $B$ are $2\delta^2$, and all of the powers of $x$ and $y$ exist, 
the polynomials quickly become very cumbersome, so we will not try to 
give even all of the four models given by Tomimatsu and Sato.  For $\delta=1$
we have 
 \begin{equation}
A= p^2 (x^2 -1) - q^2 (1-y^2),
\end{equation}
\begin{equation}
B= (px+1)^2 + q^2 y^2 ,
\end{equation}
\begin{equation}
C=-px -1.
\end{equation}
This is just the Kerr metric already studied in Ref. \cite{prd}.  The only 
model we will study in detail is the $\delta=2$ model, where
\begin{equation}
A= [p^2 (1-x^2)^2 +q^2 (1-y^2)^2]^2 + 4p^2 q^2 (1-x^2) (1-y^2) (x^2-y^2)^2 ,
\label{aeq1}
\end{equation}
\bea
B &=& \{p^2 (1+x^2) (1-x^2)+q^2 (1+y^2) (1-y^2) +2px (1-x^2)\}^2
\nonumber \\
&&+4 q^2 y^2 \{px (1-x^2)-(px+1) (1-y^2)\}^2, \label{beq1}
\eea
\bea
C &=& p^3x (1-x^2) \{-2 (1+x^2) (1-x^2) + (x^3+3) (1-y^2)\}
\nonumber \\
&&+ p^2 (1-x^2)\{-4x^2 (1-x^2)+ (3x^2+1) (1-y^2)\} +q^2 (px+1) (1-y^2)^3.
\label{ceq}
\eea
For $x$ = cos$(e^{-\tau})$, $y$ = cos $\theta$, we have
\bea
A &=& p^4 {\rm sin}^8 e^{-\tau} + q^4 {\rm sin}^8\theta + 2p^2 q^2 {\rm sin}^2e^{-\tau} {\rm sin}^2 \theta
\{ 2 {\rm sin}^4 e^{-\tau}
\nonumber \\
&&+ 2 {\rm sin}^4 \theta - 3 {\rm sin}^2 e^{-\tau} {\rm sin}^2 \theta \} 
\label{aeq2}
\eea
\bea
B &=& \{p^2 {\rm sin}^2 e^{-\tau} (1+\cos^2 e^{-\tau})+q^2 {\rm sin}^2 \theta 
(1+\cos^2 \theta ) +2p\cos e^{-\tau} {\rm sin}^2 e^{-\tau} \}^2
\nonumber \\
&&+4q^2 \cos^2\theta \{p \cos e^{-\tau} {\rm sin}^2 e^{-\tau} - (p\cos e^{-\tau}
+1) {\rm sin}^2 \theta \}^2,
\label{beq2}
\eea
\bea
C &=& p^3 \cos e^{-\tau} {\rm sin}^2 e^{-\tau} \{-2 (1+\cos^2 e^{-\tau}) {\rm sin}^2 e^{-\tau}
+ (3+\cos^3 e^{-\tau}) {\rm sin}^2 \theta \}
\nonumber \\
&&+p^2 {\rm sin}^2 e^{-\tau} \{-4\cos^2 e^{-\tau} {\rm sin}^2 e^{-\tau} + (3\cos^2 e^{-\tau}
+1) {\rm sin}^2 \theta \}
\nonumber \\
&&+ q^2 (p\cos e^{-\tau} +1) {\rm sin}^6 \theta .
\label{ceq2}
\eea
These expressions can now be used in (\ref{lambda}) and (\ref{pandq}) 
to give $P$, $Q$ and 
$e^{-\lambda/2} e^{\tau/2}$.  In Figures (1) and (2) we give the behavior
of $P -m  = \ln (\frac{|A|}{|B|})$ [i.e. $m = -\ln(\sin e^{-\tau})$] and $Q=2(q/p)
{\rm sin}^2 \theta(C/A)$as functions of $\theta$ for several values of $\tau$.

In Ref. \cite{prd} the behavior of $P$ was that of a single peak in $P$ as a 
function of $\theta$ that varied in width and height as $\tau$ 
varied from  $-{\rm ln} \pi$ to $+ \infty$. 
In the $\delta=2$ case we can see from Figs. (1 a-g) that there are two new
peaks that move in $\theta$ as $\tau$ advances (near $\tau = -\ln \pi$ and
as $\tau \rightarrow \infty$ the peaks are too small to appear in the
figures).  Figures (2 a-g) show the behavior of $Q$ as a function of $\theta$
for the same values of $\tau$ as in Fig. (1).  The same two new peaks appear
in $Q$.  The flat lines at $\tau = -\ln \pi$ and $\tau = \infty$ can be seen as
square peaks where the extra peaks have been ``squashed'' into the vertical
axes at $\theta = 0$ and $\theta = \pi$.  Since the Tomimatsu-Sato polynomials
$A$ and $B$ have powers up to $2\delta^2$, we can expect similar behavior
for higher $\delta$ models, with a rapidly growing number of peaks as $\delta$
grows.  Since the amplitude of $P - m$ varies considerably, we have given
$P - m$ for $\theta = 0$ as a function of $\tau$ in Figure (3).

There are a number of points about $A$ and $B$.  For $\delta =1$,
$e^{\lambda/2}
e^{\tau/2}$ has no coordinate singularity at $\cos^2 e^{-\tau} = \cos^2 \theta$,
since $e^{2\gamma}$ has a factor of $[\cos^2 e^{-\tau}$ $-\cos^2 \theta]^{-1}$ 
that cancels the same factor in Eq. (\ref{lameq}).  For all other values of $\delta$ 
there is a coordinate singularity where $\cos^2 e^{-\tau}$ = $\cos^2 \theta$.
Another problem is the possible existence of singularities (which can 
be either coordinate singularities or curvature singularities) corresponding 
to points where $A$ or $B$ are zero.
  
We can check whether either of these polynomials is zero. If we define 
$V= {\rm sin}^2 e^{-\tau}$ and $W = {\rm sin}^2 \theta$, we find that 

\begin{equation}
A= (p^2 V^2 + q^2 W^2)^2 +4p^2 q^2 VW (W-V)^2.
\end{equation}

Notice that this expression is the sum of two positive terms 
$(V,W \geq 0)$ so the second term in $A$ can only be zero if $W$ or $V$ 
is zero (the first term being zero then implies $W =V =0$) or if $W=V$ and 
$(p^2 + q^2) V^2 = V^2 =0$.  Both of these conditions imply that the only 
zeros of $A$ occur at $\cos e^{-\tau} = \pm 1$.  Tomimatsu and Sato \cite{ts}
and Hikida and Kodama \cite{jps} have shown that these singularities are only
the inner and outer horizons of the model.
 
The other possibility of singular points is where $B=0$.  From (\ref{beq2}) 
we can see that $B$ is also the sum of two positive terms, and these terms must
both be zero.  There are various possibilities.  The second term is zero if 
$\cos \theta =0$ or $q=0$ $(p=1)$ or 

\begin{equation}
p \cos e^{-\tau} {\rm sin}^2 e^{-\tau} - (p \cos e^{-\tau} +1) {\rm sin}^2 \theta = 0 
\label{b0eq}
\end{equation}
For $\cos \theta =0$, the first term in $B$ reduces to (using {\rm sin}$^2 e^{-\tau} =1
- \cos^2 e^{-\tau}$ and $p^2+q^2 = 1)$
\begin{equation}
p^2 \cos^4 e^{-\tau}+ 2p \cos^3 e^{-\tau} -2p \cos e^{-\tau} - 1 =0.
\end{equation}

The four solutions of this equation are real only for $|p| > 1$ (impossible)
except for two cases, $p=+1$, $\cos e^{-\tau} = +1$, $p = - 1$, $\cos e^{-\tau} = 1$.
These are ``ring" singularities on the horizons of the Gowdy models.  This is 
the limit of the well-known ring singularity of the Tomimatsu-Sato 
$\delta =2$ metrics that is always outside the Gowdy region except for the case 
we have given.  This singularity is a curvature singularity \cite{ts,jps}.

The next possibility is $q=0$ $(p = \pm 1)$ where the first term in $B$ 
is zero if ${\rm sin} e^{-\tau} = \pm 1$ or if $\cos^2 e^{-\tau} + 2 \cos e^{-\tau}
+1 =0$ (which has a real solution only for $\cos e^{-\tau}=-1$).  These are 
the horizons of the Tomimatsu-Sato metric, and the curvature invariants
are finite except for the ring singularity when $p=\pm 1$.
 
Finally, we can take any solution where (\ref{b0eq}) is satisfied.  Solving 
for sin$^2 \theta$ and plugging the result into the first term 
of $B$, we find a fifth order expression in cos $e^{-\tau}$ that must
be zero.  Plotting this expression as a function of $e^{- \tau}$ for $p$ 
between  0 and 1, we find that it is never zero except for $p=\pm 1$ 
for $\cos e^{-\tau} = +1$.  Each of these gives ${\rm sin}^2 \theta =0$, so 
the singularity is just the ordinary coordinate singularity of spherical 
coordinates at the poles.

\section{THE EREZ-ROSEN SOLUTION}

The Erez-Rosen solution \cite{erro} belongs to the class of static 
solutions and its metrics functions read ($q$ is an arbitrary
constant)
\begin{equation}
\omega = 0,  
\end{equation}
\begin{equation}
f = {\rm exp} (2\psi ), \quad \psi = \frac{1}{2} \ln
\frac{x-1}{x+1} + q\tilde\psi, 
\label{psi}
\end{equation}
\begin{equation}
\gamma = \frac{1}{2} \ln \frac{x^2-1}{x^2-y^2} +q\tilde\gamma, 
\label{gamma}
\end{equation}
where
\begin{equation}
\tilde\psi = \frac{1}{2} (3y^2 -1) \left[\frac{1}{4} (3x^2-1) \ln
\frac{x-1}{x+1} + \frac{3}{2} x \right],
\label{psir}
\end{equation}
and 
\begin{eqnarray}
\tilde\gamma &=& \left( 1+ \frac{1}{2} q\right) \ln \frac{x^2 -1}{x^2 -y^2}-
\frac{3}{2}(1-y^2) \left(x\ln \frac{x-1}{x+1} +2 \right) \nonumber \\
&+& \frac{9}{16}q (1-y^2) \left[ x^2 + 4y^2 - 9x^2 y^2 - 4/3 +x (x^2 +7y^2 - 9
x^2 y^2 -5/3) \ln \frac{x-1}{x+1} \right. \nonumber \\
& & \left. + \frac{1}{4} (x^2 -1) (x^2+y^2 -9 x^2y^2 -1) \ln^2
\frac{x-1}{x+1} \right].
\end{eqnarray}

The Erez-Rosen solution is interpreted as describing 
the exterior gravitational field
of a non-spherically symmetric mass distribution with 
quadrupole moment 
proportional to the arbitrary constant $q$. 
When the mass quadrupole moment
vanishes we recover the standard Schwarzschild spacetime with $x=r/M -1$
and $y=\cos\theta$. In general, for the metric functions of the Erez-Rosen
solution to be well defined one has to demand that $|x|>1$ (and $|y|\leq 1$), 
although the limiting value $x=\pm 1$ can also be included by using an
appropriate procedure \cite{fort}. For the purposes of the present work,
however,
we will include that limiting value in the time dependent sector,
which now corresponds to $-1\leq x \leq 1$. In this case, there is no
problem with the argument of the logarithmic function appearing in the 
metric functions (\ref{psi}) and (\ref{gamma}). In fact, one can show
that the argument $(x-1)/(x+1)$ can be replaced everywhere by its absolute 
value and the resulting expressions remain valid as exact solutions of the 
vacuum field equations.

In order to identify this solution as an $S^1 \times S^2$ Gowdy model 
in the region $-1 \leq x \leq 1$, 
we take, as before, $y= \cos \theta$ and 
$x = \cos e^{-\tau}$.  The metric functions $A$ and $B$ 
can then be chosen as 
\begin{equation}
A = - {\rm sin}^2 e^{-\tau} {\rm exp} (2q \tilde \psi ) \quad {\rm and} \quad B = (1 + \cos
e^{-\tau})^2
\end{equation}
As in the previous cases, the function $\gamma$ turns out to be 
proportional to $A$ since  ${\rm exp}(2 \gamma) =  A (\cos^2 e^{- \tau}
- \cos^2 \theta)^{-1} {\rm exp}(2 q\tilde\gamma - 2 q \tilde \psi )$.
Consequently, the metric functions for the corresponding Gowdy model 
can be written as 
\begin{equation}
Q = 0
\end{equation}
\begin{equation}
e^P = \frac{{\rm sin} \, e^{-\tau}}{(1+\cos e^{-\tau})^2} e^{2q\tilde\psi }, 
\label{perro1}
\end{equation}

\begin{equation}
e^{-\lambda /2+\tau /2} = (1+\cos e^{-\tau})^2 
e^{2q\tilde\gamma -2q\tilde \psi } . 
\end{equation}
We now have a time and angular dependent background which evolves in time
from $\tau=-\ln\pi$ to $\tau\rightarrow \infty$. 

Since $Q = 0$ this is a polarized Gowdy model, so there is a separable
general solution for $P$ as a sum over Legendre polynomials,
\begin{equation}
P = \sum^{\infty}_{l = 0} T_{l} (e^{-\tau}) P_l (\cos \theta).
\label{perro}
\end{equation}
Erez-Rosen is a particular solution where the expression (\ref{perro1}),
using (\ref{psir}), reduces to
\begin{equation}
P = -2Q_0 (\cos e^{-\tau}) P_0 (\cos \theta) - 2qQ_2(\cos e^{-\tau})
P_2 (\cos \theta),
\end{equation}
where the $Q_l$ are Legendre functions of the second kind.  This is a
relatively simple function, and if we use $P_0 = 1$ to write
\begin{equation}
P + 2Q_0 (\cos e^{-\tau}) = -2qQ_2(\cos e^{-\tau}) P_2 (\cos \theta),
\end{equation}
$P + 2Q_0$ is simply a time-dependent amplitude times $P_2 (\cos \theta)$,
a very well known function.  For completeness, however, we will graph $P_2
(\cos \theta)$ versus $\theta$ in Figure (4) and $Q_2 (\cos e^{-\tau})$
($1/q$ times the amplitude of $P + 2Q_0$ at $\theta = 0$) versus $\tau$
in Figure (5).  Notice that $Q_2 (\cos e^{-\tau})$ is just $Q_2 (\cos z)$
``stretched'' by the exponential factor $e^{-\tau}$.

One can calculate the
Kretschman scalar corresponding to this spacetime and show that 
no curvature singularities exist in the region $-\ln\pi < \tau < \infty$.
However, the limiting values of this region show interesting behavior.
To see this, let us turn back to the Erez-Rosen metric since we know
that these limiting values of the cosmological evolution correspond 
to the horizon  of the exterior ``black hole'' solution. 
The horizon in the Erez-Rosen metric is determined by the zeros 
 of the norm of the Killing vector $\xi^\mu =(\partial_t)^\mu$ 
associated with the time coordinate $t$:
\be
\xi^\mu \xi_\mu = { \exp[3q x P_2(y)] \over (x+1)^{1+qP_2(x) P_2(y)} } 
(x-1)^{1+qP_2(x) P_2(y)}\ .
\ee
Here the horizon turns out to be at $x=+1$ with 
the interesting feature that its existence depends on the 
values of the constant $q$ and the angular coordinate $y$.  In fact, 
the hypersurface $x=+1$ is a horizon only if the condition
\begin{equation}
1+ \frac{q}{2} (3y^2-1) > 0 ,
\end{equation}
is satisfied, i.e. for
\begin{equation}
q>0  \quad {\rm and} \quad \frac{1}{3} (1-2/q)\leq y^2 \leq 1 ,
\end{equation}
\begin{equation}
q<0 \quad {\rm and} \quad 0 \leq y^2 \leq \frac{1}{3} (1-2/q) .
\end{equation}

Notice that when these conditions are not satisfied the norm of 
the Killing vector diverges (Killing singularity). 
An analysis of these conditions leads to the following conclusions:
For $-1\leq q\leq 2$ the horizon occupies the entire hypersurface $x=1$,
i.e. it does not depend on the angular coordinate $y$. This, of course,
includes the limiting case $q=0$ (the Schwarzschild limit) in which
no Killing singularity exists at $x=1$. 
If we represent the
hypersurface $x=1$ by a circle, the horizon coincides with the entire 
circle.
For $q< -1$, the horizon is symmetric with respect to the equatorial
plane $y=0$ ($\theta = \pi/2$), but it does not cover the entire
hypersurface $x=1$. Indeed, around the symmetry axis $(y=\pm 1)$ a 
Killing singularity appears that extends from $\theta = 0$ to 
$\theta = \theta_- ={\rm arccos}\sqrt{(1+2/|q|)/3}$. The 
arc-length of the section occupied by the Killing singularity reaches
its maximum value when $q\rightarrow -\infty$, i.e., for $\theta_{sing} =
{\rm arccos}\sqrt{1/3}$. 
For positive values of $q$ and $q>2$, the horizon at $x=1$ is 
symmetric with respect to the symmetry axis ($y=\pm 1)$ and reaches
its maximum arc-length with respect to the axis at 
$\theta_{hor} = \pi/2 - {\rm arccos}(1/\sqrt{3})$ for 
$q\rightarrow \infty$. The remaining section of the hypersurface $x=1$ 
is covered by the Killing singularity.
There is a particular angular direction
\be
\theta_- = {\rm arccos}\sqrt{{1\over 3}\left( 1 + {2\over |q|}\right)}
\ , \qquad {\rm for} \qquad q < -1 \ ,
\label{angn}
\ee
or 
\be
\theta_+ = {\rm arccos}\sqrt{{1\over 3}\left( 1 - {2\over q}\right)}
\ , \qquad {\rm for} \qquad q > 2 \ ,
\label{angp}
\ee
which determines the boundary on $x=1$ that separates the horizon from
the Killing singularity. In fact, due to the axial symmetry this 
boundary corresponds to a sphere $S^1$. So we see that in the case
$q<-1$ and $ q > 2$, the hypersurface $x=1$ contains a horizon and
a Killing singularity as well. The corresponding Gowdy cosmological 
model is defined ``inside'' this hypersurface. 

Since the analytic expression for the Kretschman scalar in this case
is rather cumbersome, we only quote the results of our analysis. 
There exists a true curvature singularity at $x=-1$, independent of
the values of $q$ and the angular variable $y$. (This corresponds to 
the Schwarzschild singularity at the origin of coordinates.)
 A second singularity
is situated on the hypersurface $x=1$ for all values of $q$ and $y$,
except on the symmetry axis $(\theta = 0)$ and for the special 
direction $\theta_-$ (or $\theta_+)$, i.e., on the boundary 
between the ``horizon'' and the Killing singularity. In those particular
directions the Kretschman scalar remains constant. 
This means that the cosmological model inside the ``singular horizon''
evolves from a true Big Bang singularity at $x=-1$ ($\tau = -\ln\pi$)
into a true Big Crunch singularity everywhere at $x=1$ ($\tau \rightarrow
\infty$), except in the special directions 
$\theta =0, \ \theta_-, \theta_+$. From the point of view of the exterior
($|x|>1$) ``black hole'' spacetime, the interior $(|x|\leq 1)$ cosmological
model can be reached only through the ``angular windows'' located
at $\theta =0, \ \theta_-, \theta_+$.

\section{CONCLUSIONS}

Time-dependent backgrounds can be generated in a systematic way by means of
what was
called  ``horizon methods" in \cite{samos}. 
In \cite{prd}  we applied this procedure to the Kerr 
metric and  were able to  obtain a time and ``angular" dependent background, 
a Gowdy $S^1 \times S^2$ model. In this work we have extended the 
previous analysis and have been able to exhibit backgrounds depending on time 
and an ``angular" variable, Gowdy $S^1 \times S^2$ models that correspond to 
each of the  three categories of axisymmetric solutions, the simple Zipoy-Voorhees 
metric, Tomimatsu-Sato metrics and the 
Erez-Rosen metric which has complicated curvature-singularity-horizon
behavior.

In the case of the Zipoy-Voorhees metric we have shown that the outer and inner
``horizons" are actually surfaces of infinite curvature (naked singularities) 
and the models are then untenable as black hole models because they violate
cosmic censorship. However, between the ``horizons" it represents a viable
time and ``angular" dependent background.

The Gowdy models generated by the Tomimatsu-Sato metrics we studied (including
the Kerr metric of Ref. \cite{prd}) are unique
among our three metrics in that they are unpolarized Gowdy models ($Q \neq 0$).
This means that it is the only solution that cannot be written as a
linear sum over eigenfunctions (Legendre polynomials).  Another point is
that the cosmological singularities are just horizons except for the
extreme $\delta = 2$ model with $p = \pm 1$ and $q = 0$, where there is a
ring curvature singularity at $\theta = \pi/2$ on the horizons.

In the case of the Erez-Rosen metric we have seen that it can be interpreted
as a polarized Gowdy model inside the ``singular horizon''. This horizon is special 
in the sense that for a large range of values of the ``quadrupole moment''
it allows the existence of a ``regular horizon'' (with an $S^1$ topology)
through which the interior time and angular dependent sector can be reached. 

The Tomimatsu-Sato metrics and the Erez-Rosen metric represent
long-wavelength unpolarized and polarized ``gravitational waves'' propagating
around the universe (the Zipoy-Voorhees metric is a degenerate example that
does not depend on the ``angular'' variable).  The field-theory $S$-brane
solutions generated from these metrics may be of interest as simple
field-theory brane models.

S-brane solutions have been generated by means of the Kerr metric.
These solutions have been proposed and studied independently 
in \cite{tasi} and \cite{wang}.
We have outlined the possibility of using the procedure presented
here and in \cite{prd} to generate models that would correspond to 
generalized $S$-brane solutions.
The corresponding time dependent solutions to the three different 
categories of axisymmetric solutions presented here also depend on an 
``angular"  coordinate.   An analysis of the  models 
that one would obtain by introducing an $i$-factor in the coordinates 
is in progress and the S-brane solutions generated by this procedure 
will be discussed elsewhere.

\section*{ACKNOWLEDGMENTS}

This work was supported in part  by DGAPA-UNAM grant IN112401,  
CONACyT-Mexico grant 36581-E, and US DOE grant DE-FG03-91ER 40674.
H.Q. thanks UC MEXUS-CONACyT (Sabbatical Fellowship Programm) 
for support.

\section*{APPENDIX}

As we have mentioned several times, the only obvious singularity in the three
Gowdy metrics we have studied is the one where $\cos^2 e^{-\tau} = \cos^2
\theta$ ($x^2 = y^2$ in Lewis-Papapetrou coordinates) which makes
$e^{-\lambda/2 + \tau/2}$ either zero or infinity on that spacetime surface.
One could check that curvature invariants are nonsingular on that surface,
but for our metrics it is simple enough to find a coordinate transformation
that makes the metric nonsingular.

In the $\tau \theta$ sector of the Gowdy metric we can write the two-dimensional
metric, $d\sigma^2 = e^{-\lambda/2}e^{\tau/2}(-e^{-2\tau} d\tau^2 +
d\theta^2)$, and for our metrics we have
\begin{equation}
d\sigma^2 = \frac{W(\tau, \theta)}{|\cos^2 e^{-\tau} - \cos^2 \theta|^s}
(-e^{-2\tau} d\tau^2 + d\theta^2), \label{dsig}
\end{equation}
where the power $s$ may be positive or negative, and $W(\tau, \theta)$ is
nonsingular for $\tau \neq -\ln \pi$ or $\tau \neq \infty$.

For the Zipoy-Voorhees metric,
\begin{equation}
W(\tau, \theta) = \frac{(\sin^2 e^{-\tau})^{\delta^2 + \delta}}{(
\cos e^{-\tau}-1)^{2\delta}} \label{wzip}
\end{equation}
and $s = \delta^2 - 1$.  Notice that $s$ is negative for $-1 < \delta <
+1$ and positive otherwise, and $W$ is nonsingular and non-zero except at
$\tau = -\ln \pi$ or $\tau = \infty$.

For any Tomimatsu-Sato metric,
\begin{equation}
W(\tau, \theta) = \frac{1}{p^{2\delta}}B(\tau, \theta), \label{btom}
\end{equation}
$B$ a polynomial of order $2\delta^2$.  The parameter $s$ is $\delta^2 -
1$ and is positive for all $\delta > 1$.  The function $B$ is always of
the form $(\alpha + \beta)(\alpha^{*} + \beta^{*})$, $\alpha$ and $\beta$
complex functions of $\tau$ and $\theta$, so $B$ is always positive and is
zero only if $|\alpha + \beta| = 0$.  For $\delta = 2$ we showed that this
expression is zero only for special values of $p$ and $q$ given in Sec. III
and only for $\tau = -\ln \pi$ and $\tau = \infty$.  We can conjecture that
this will also be true for any value of $\delta$.

Finally, for the Erez-Rosen metric we have
\begin{equation}
W(\tau, \theta) = (1+\cos e^{-\tau})^2 (\sin^2 e^{-\tau})^{2q+q^2} 
e^{2q \gamma^* -2q\tilde\psi}, \label{wer}
\end{equation}
where
\bea
&&\gamma^* = -\frac{3}{2}\sin^2 \theta \left [ \cos e^{-\tau} \ln
\left (\frac{1 - \cos e^{-\tau}}{1 + \cos e^{-\tau}} \right )+ 2 \right ] +
\frac{9}{16}q \sin^2 \theta \bigg [ \cos^2 e^{-\tau} + 4\cos^2 \theta
\nonumber \\
&&-9\cos^2 e^{-\tau} \cos^2 \theta - \frac{4}{3} +
\cos e^{-\tau}(\cos^2 e^{-\tau} + 7 \cos^2 \theta - 9\cos^2 e^{-\tau}\cos^2\theta
\nonumber \\
&&- \frac{5}{3}) \ln \left (\frac{1 - \cos e^{-\tau}}{1 + \cos e^{-\tau}}
\right ) - \frac{1}{4} \sin^2 e^{-\tau} (-\sin^2 e^{-\tau} + \cos^2 \theta
\nonumber \\
&&- 9 \cos^2 e^{-\tau} \cos^2 \theta ) \ln^2 \left ( \frac{1 - \cos e^{-\tau}}
{1 + \cos e^{-\tau}} \right ) \bigg ], \label{gamtil}
\eea
and $s = 2q + q^2$.  Notice that $W$ is either zero or infinity only for
$\cos^2 e^{-\tau} = \pm 1$, and so the only possible singularity not on
the inner or outer horizons is where $\cos^2 e^{-\tau} = \cos^2 \theta$.
For $-2 < q < 0$, $s$ is negative.  For all other values of $q$ it is
positive.

If we now look at the general form of $d\sigma^2$ from (\ref{dsig}), and
define $\xi = e^{-\tau}$, we have
\begin{equation}
d\sigma^2 = \frac{W(\xi, \theta)}{|\cos^2 \xi - \cos^2 \theta|^s} [-d\xi^2
+ d\theta^2]
\end{equation}
\begin{equation}
= W(\xi, \theta)\left \{ \frac{-d\xi^2 + d\theta^2}{|\sin(\xi + \theta)
\sin (\theta - \xi )|^s} \right \}. 
\end{equation}
Defining $w = \xi + \theta$ and $z = \theta -\xi$,
\begin{equation}
d\sigma^2 = W(w, z) \left [ \frac{-dwdz}{|\sin w \sin z|^s} \right ].
\label{newdsig}
\end{equation}
We can now define new coordinates $u$ and $v$, where
\begin{equation}
du = \frac{dw}{|\sin w |^s}, \qquad dv = \frac{dz}{|\sin z|^s},
\label{dudv}
\end{equation}
and by carefully checking constants of integration in the different regions
where $\cos w$ and $\cos z$ have different signs, we can, in principle, find
$u$ and $v$ as functions of $w$ and $z$.  In the $uv$ coordinates we have
\begin{equation}
d\sigma^2 = -W(u,v) du dv,
\end{equation}
nonsingular as long as $W$ is neither zero nor infinity.

The integrals necessary to solve (\ref{dudv}) are not tabulated for $s$
non-integer, but if $s$ is an integer, we can express $u$ and $v$ as
finite sums of  various powers of sines of $\xi$ or $\theta$ and multiples
of $\xi$ and $\theta$ themselves for $s$ negative, and as a finite series
of sines divided by powers of cosines and logarithms of tangents for $s$
positive.  As an example, for the Tomimatsu-Sato $\delta = 2$ model given
in Sec. IV (ignoring the absolute values),
\begin{equation}
du = \frac{dw}{\sin^3 w}, \qquad dv = \frac{dz}{\sin^3 z},
\end{equation}
or
\begin{equation}
u = -\frac{1}{2} \frac{\cos(e^{-\tau} + \theta)}{\sin^2 (e^{-\tau}
      + \theta)} + \frac{1}{2} \ln \left [ \tan \left ( \frac{e^{-\tau}}{2}
      + \frac{\theta}{2} \right ) \right ]\ ,
\end{equation}
\begin{equation}
          v = -\frac{1}{2} \frac{\cos(\theta - e^{-\tau})}{\sin^2(\theta -
       e^{-\tau})} + \frac{1}{2} \ln \left [ \left | \tan \left
       (\frac{\theta}{2} - \frac{e^{-\tau}}{2} \right ) \right | \right ]
   \ .
\end{equation}

\vfill\eject
\begin{figure}[tb]
\[
\begin{array}{cc}
{\includegraphics[width = 1.50in, angle = 270]{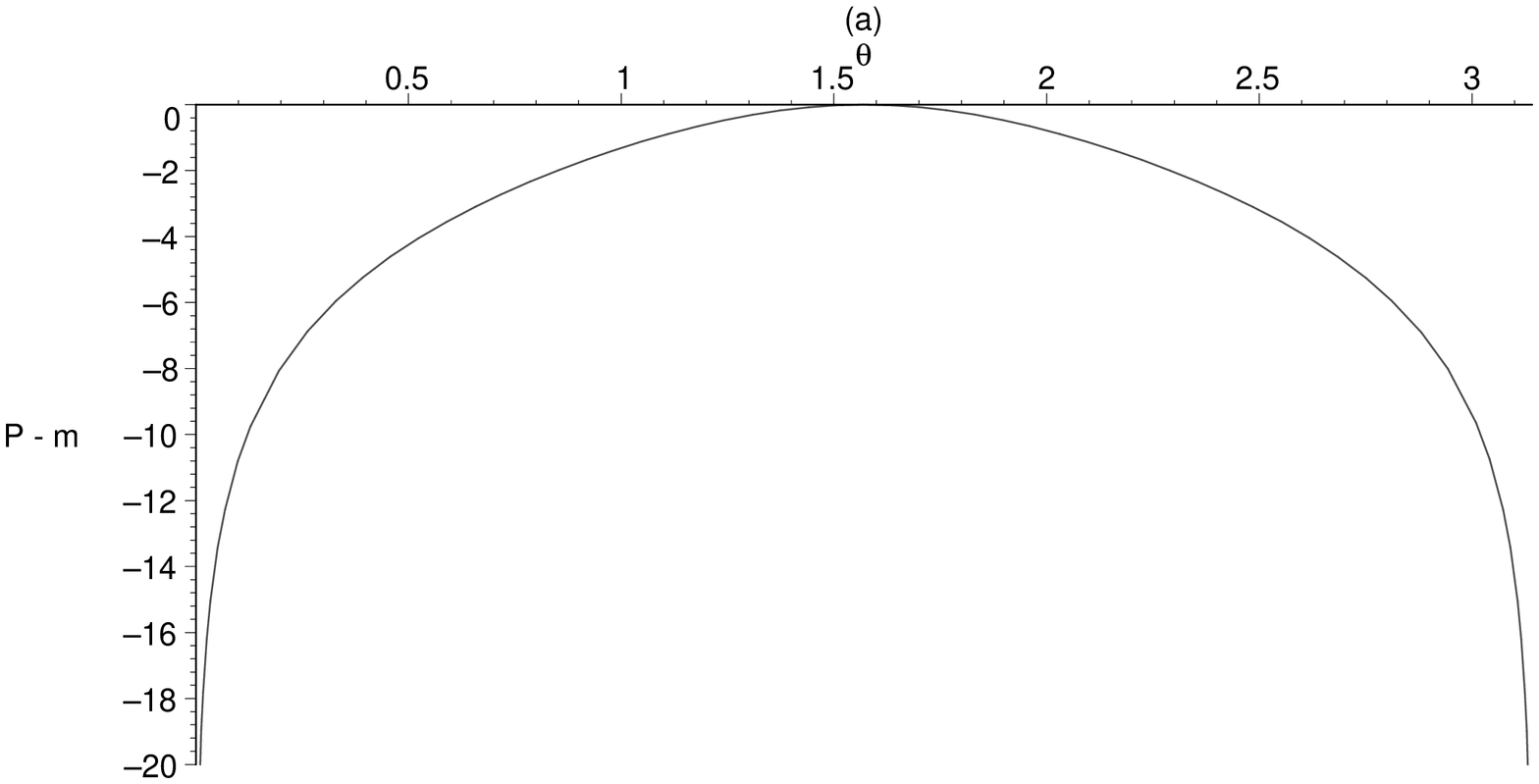}} &
{\includegraphics[width = 1.50in, angle = 270]{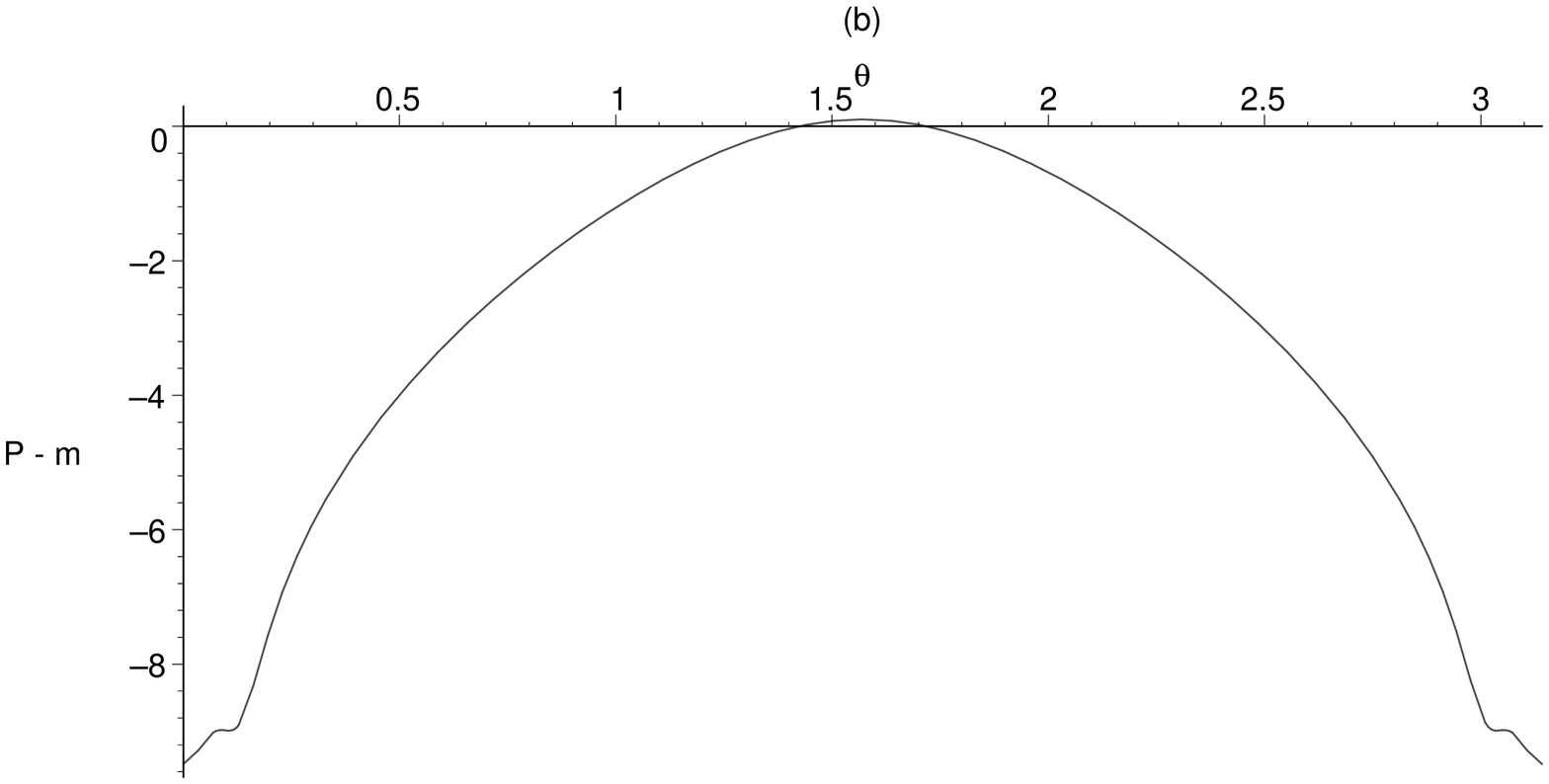}} \\
 (a)\  \tau = -\ln \pi & (b) \ \tau = -1.1 \   \\
{\includegraphics[width = 1.50in, angle = 270]{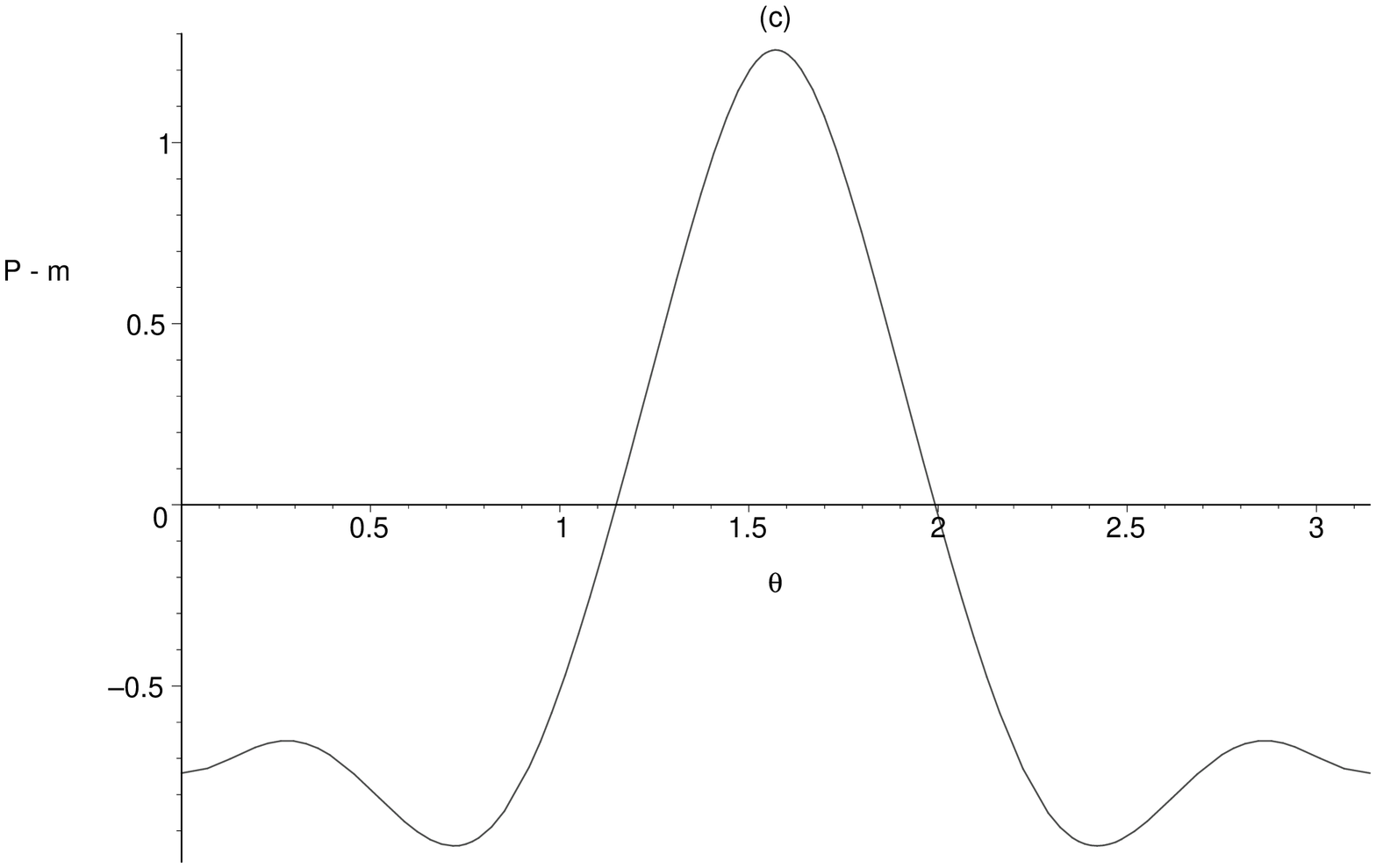}} &
{\includegraphics[width = 1.50in, angle = 270]{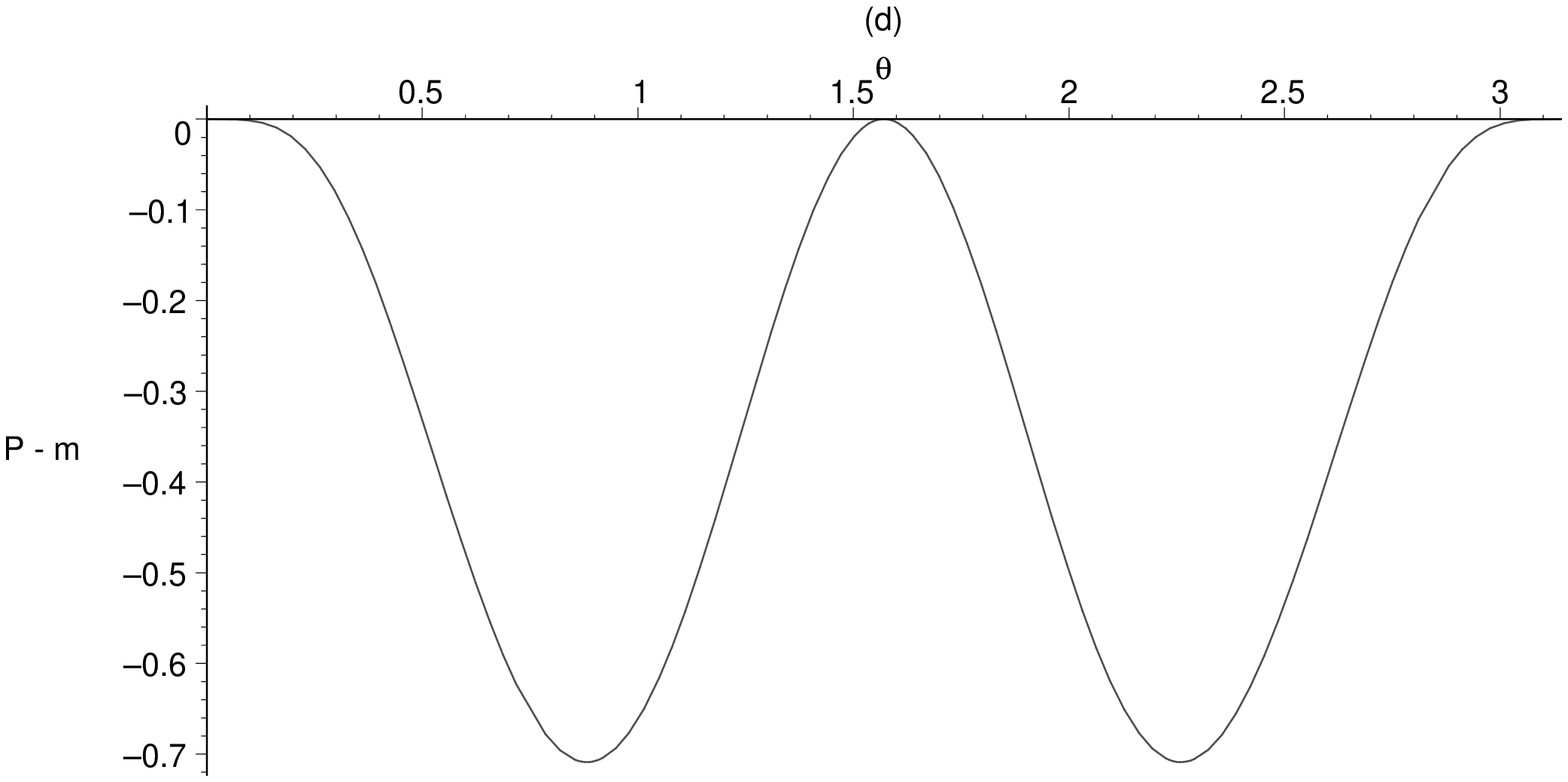}}\\
 (c) \  \tau = -0.75  &  (d) \ \tau = -\ln (\pi/2) \\
{\includegraphics[width = 1.50in, angle = 270]{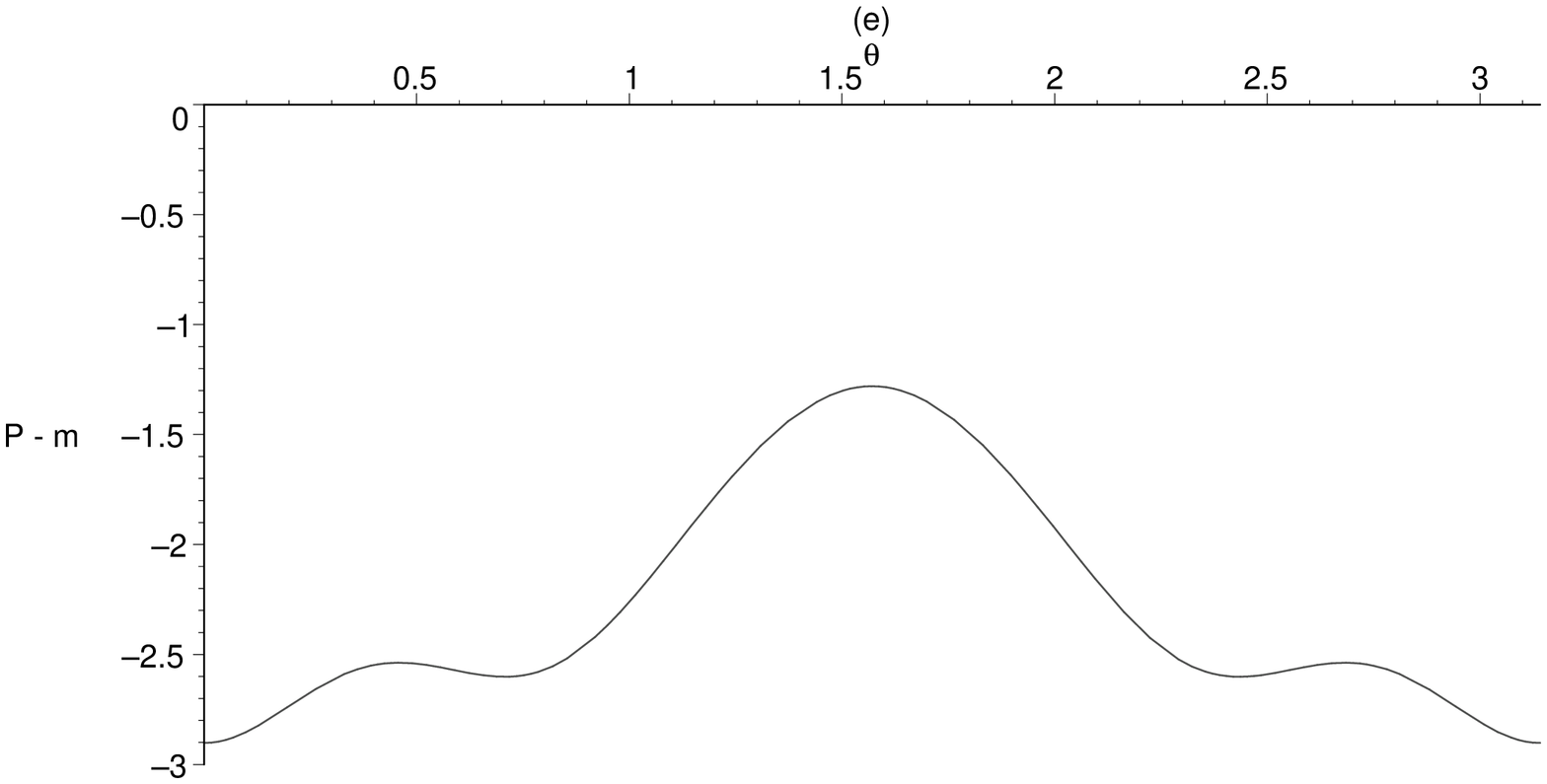}} &
{\includegraphics[width = 1.50in, angle = 270]{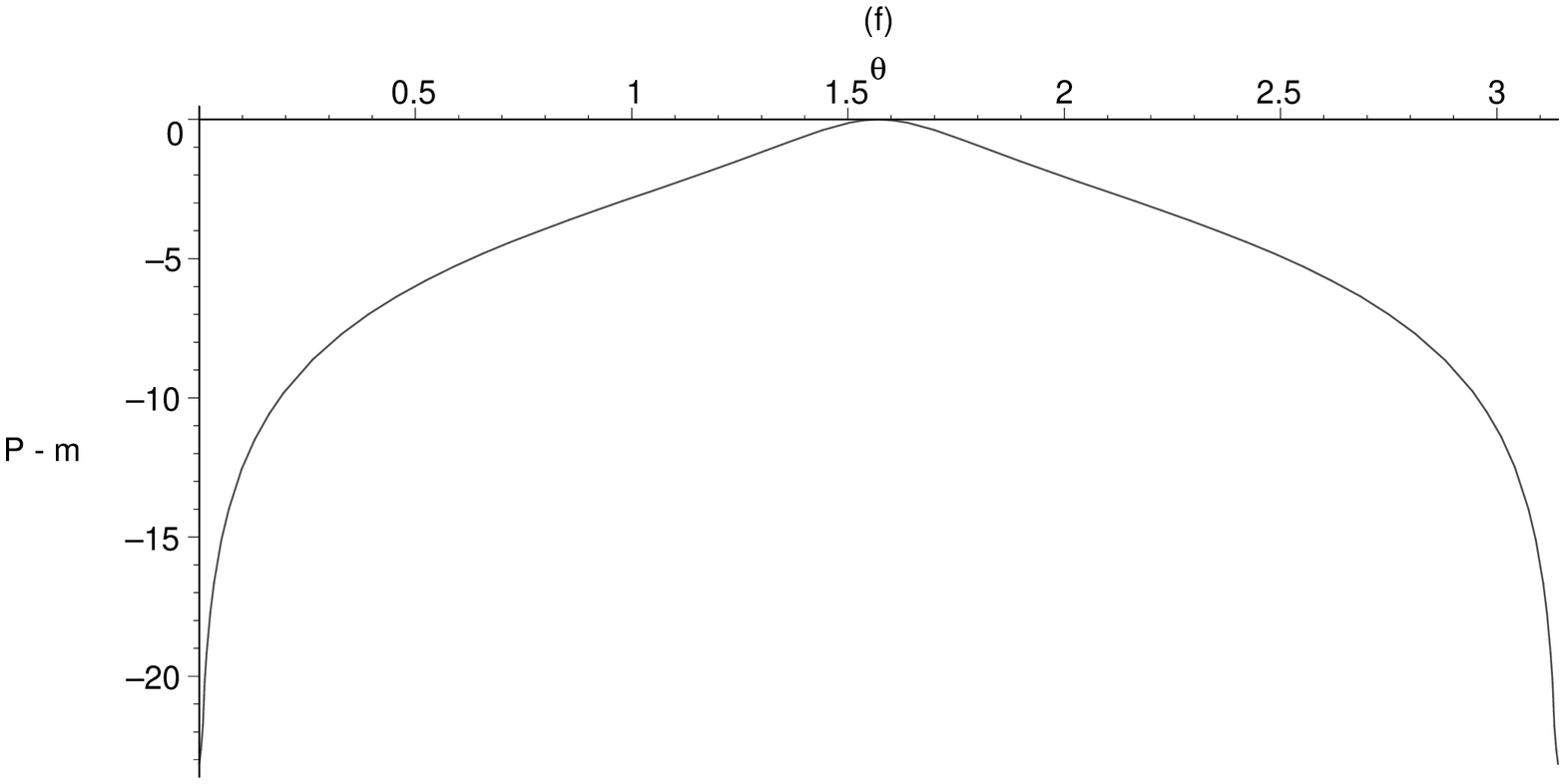}} \\
(e) \  \tau = 0 & (f)\    \tau = +5 \\
{\includegraphics[width = 1.50in, angle = 270]{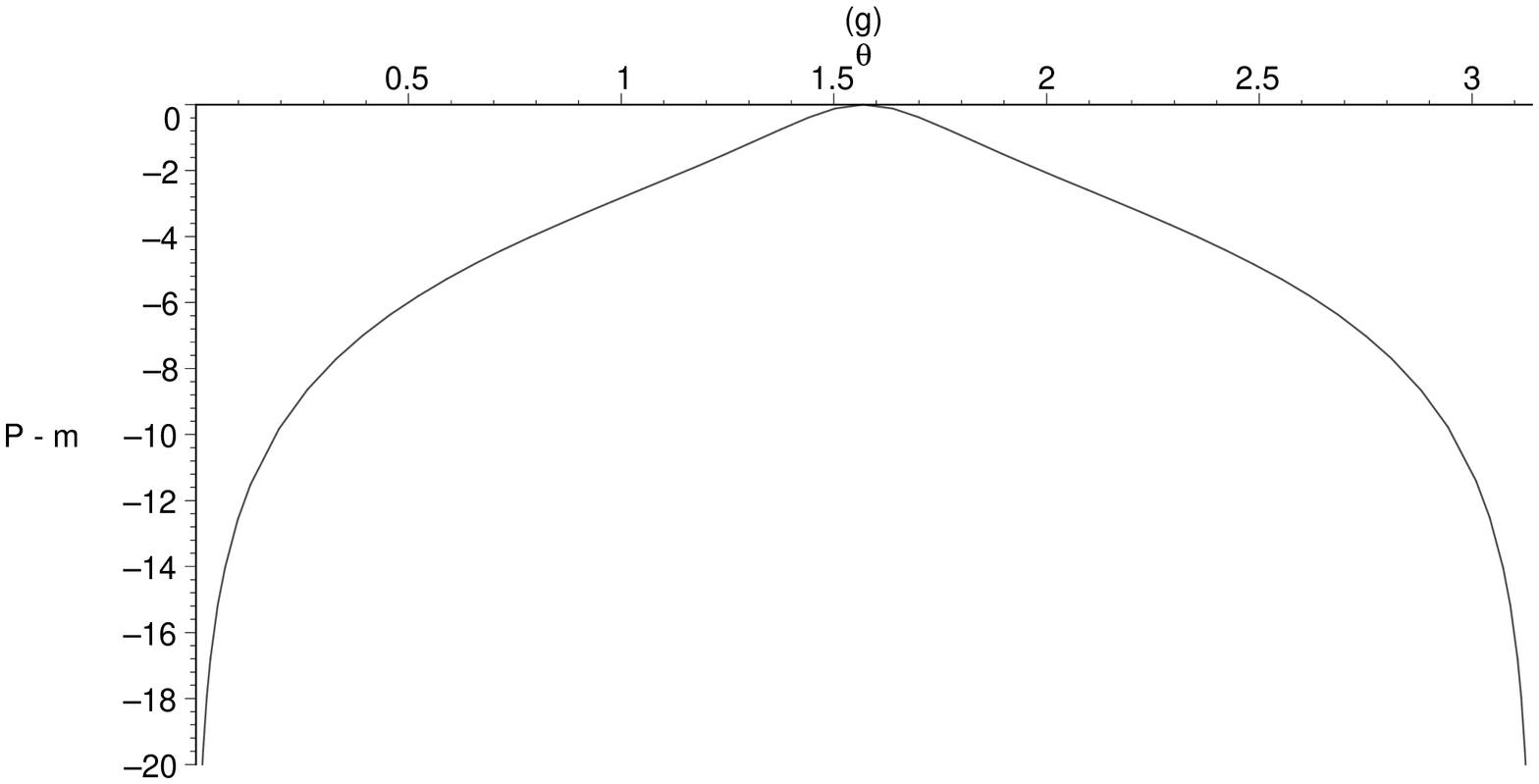}} & 
   \  \\
 (g) \ \tau = +\infty &  \  \cr
\end{array}
\] 
\caption[]{The evolution in $\tau$ of $P - m$, [$m = -\ln(\sin e^{-\tau})$], 
as a function of $\theta$. } 
\end{figure}
\vfill\eject
\begin{figure}[tb]
\[
\begin{array}{cc}
{\includegraphics[width = 1.50in, angle = 270]{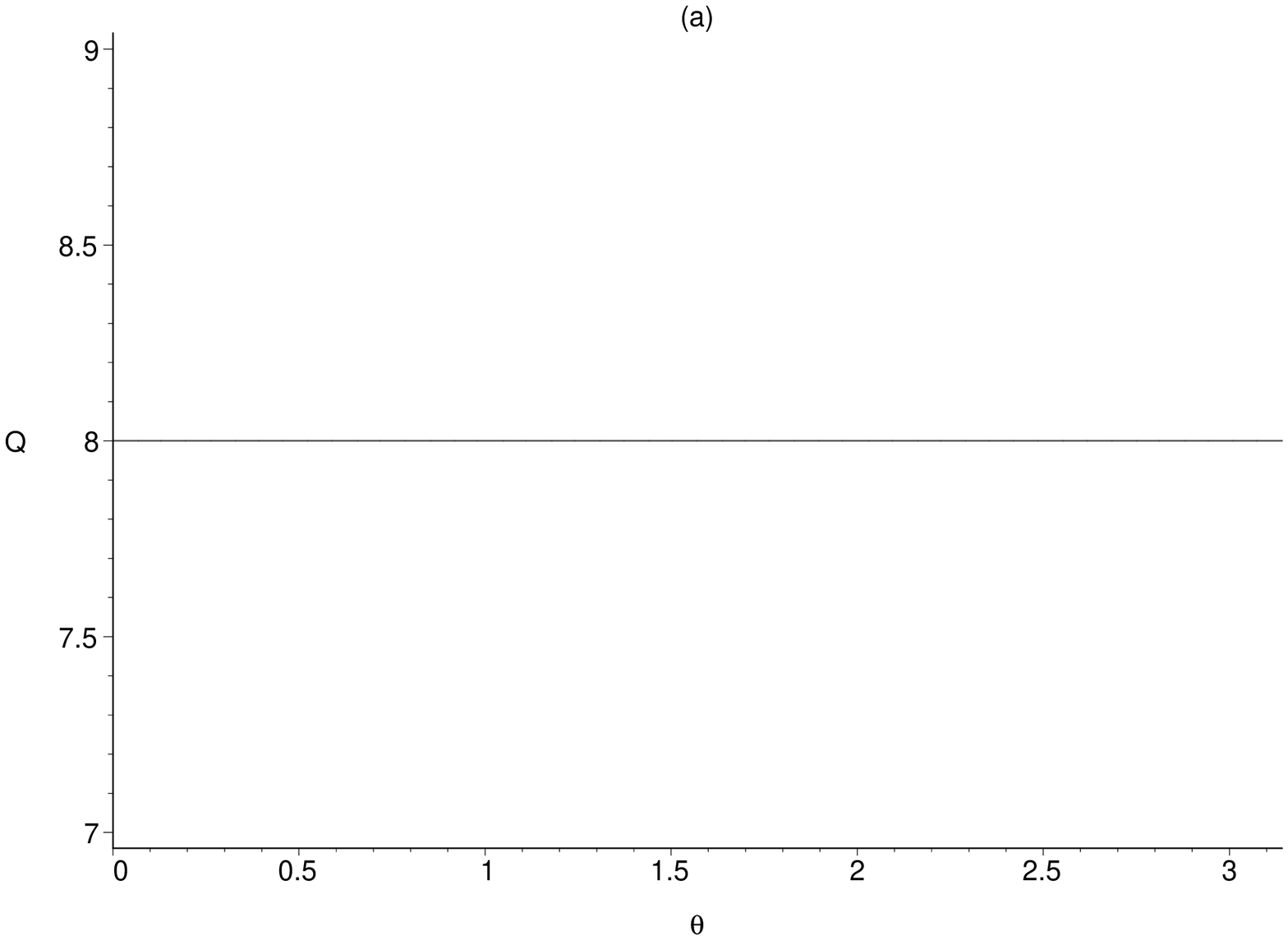}} &
{\includegraphics[width = 1.50in, angle = 270]{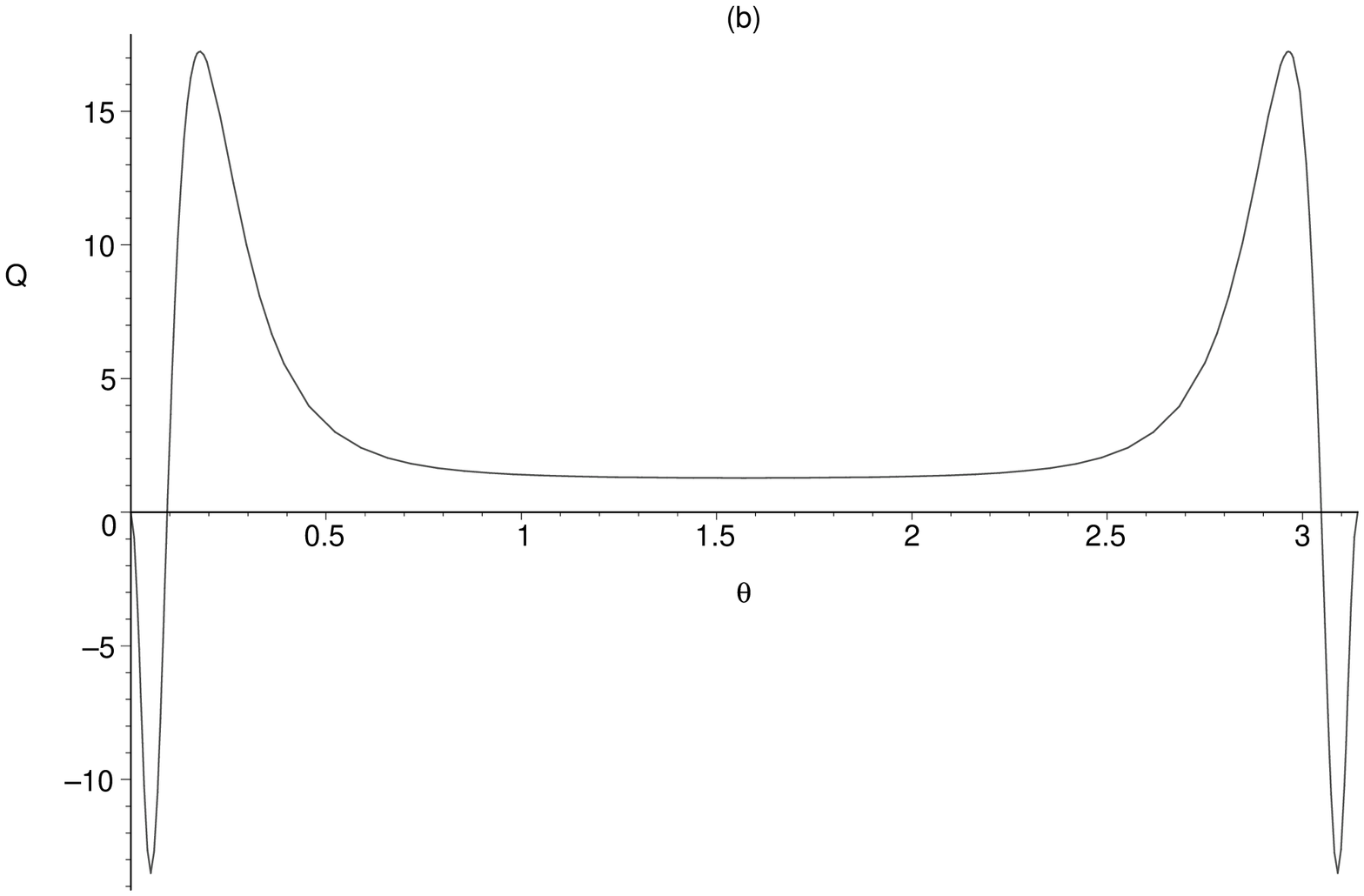}} \\
 (a)\  \tau = -\ln \pi & (b) \ \tau = -1.1 \   \\
{\includegraphics[width = 1.50in, angle = 270]{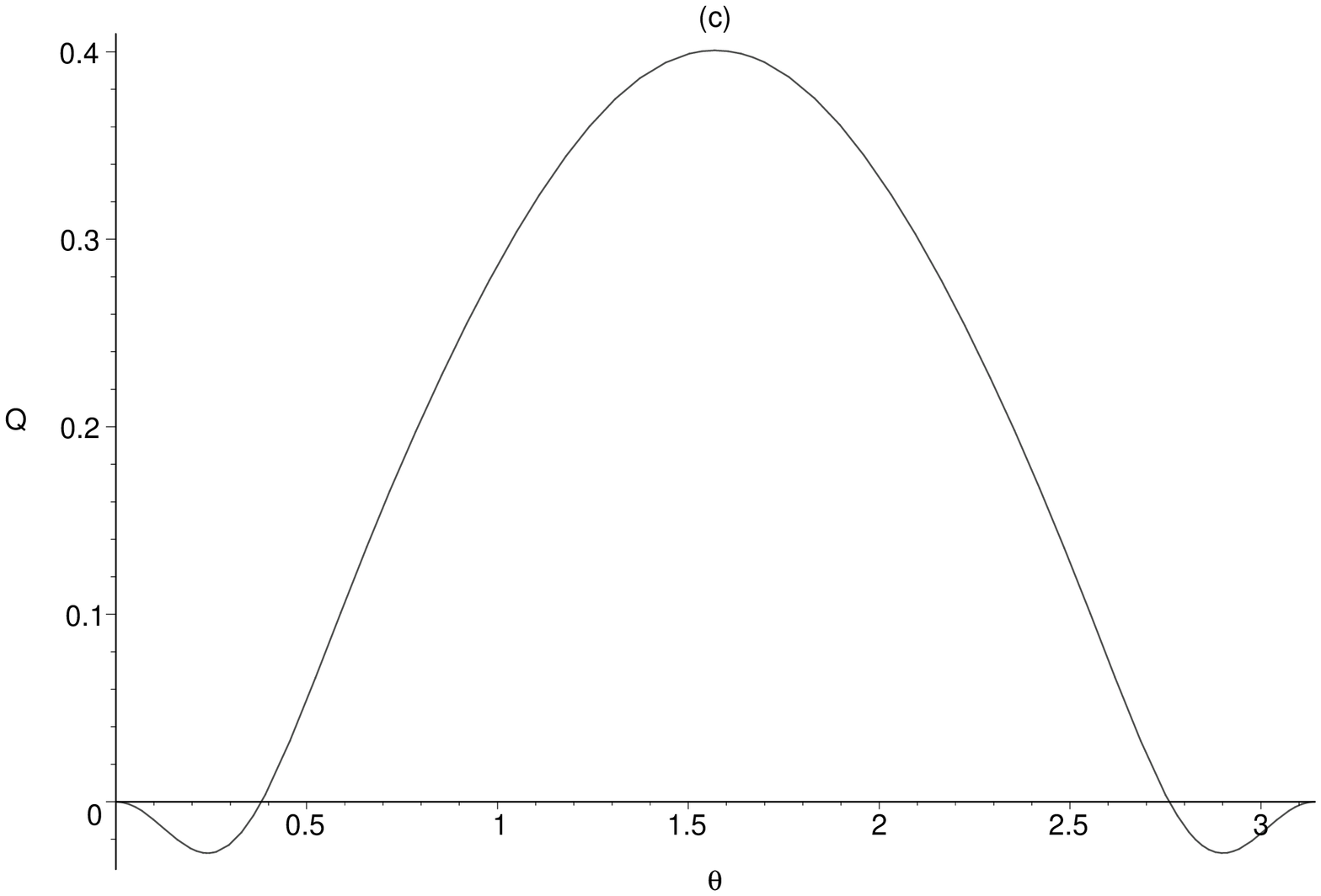}} &
{\includegraphics[width = 1.50in, angle = 270]{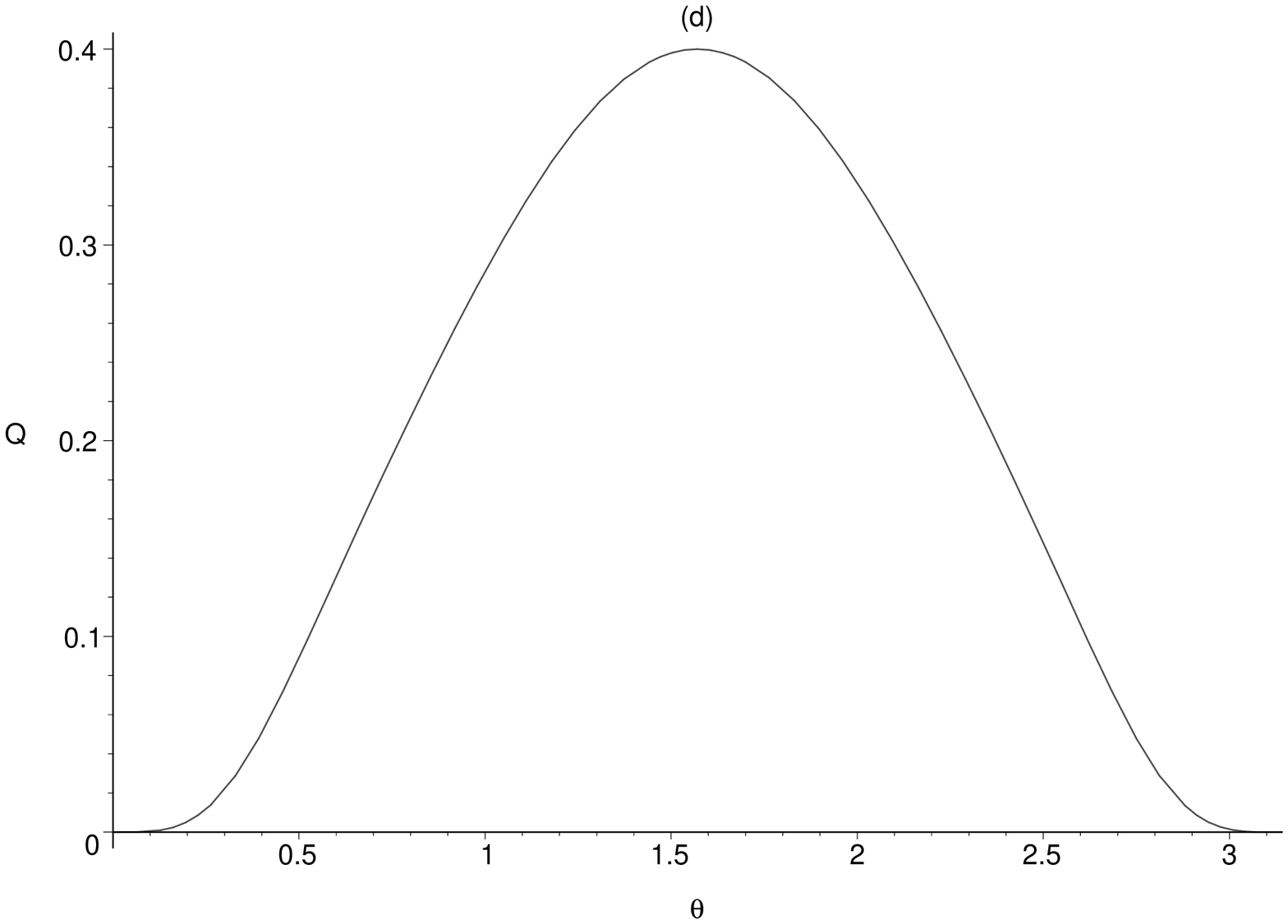}}\\
 (c) \  \tau = -0.75  &  (d) \ \tau = -\ln (\pi/2) \\
{\includegraphics[width = 1.50in, angle = 270]{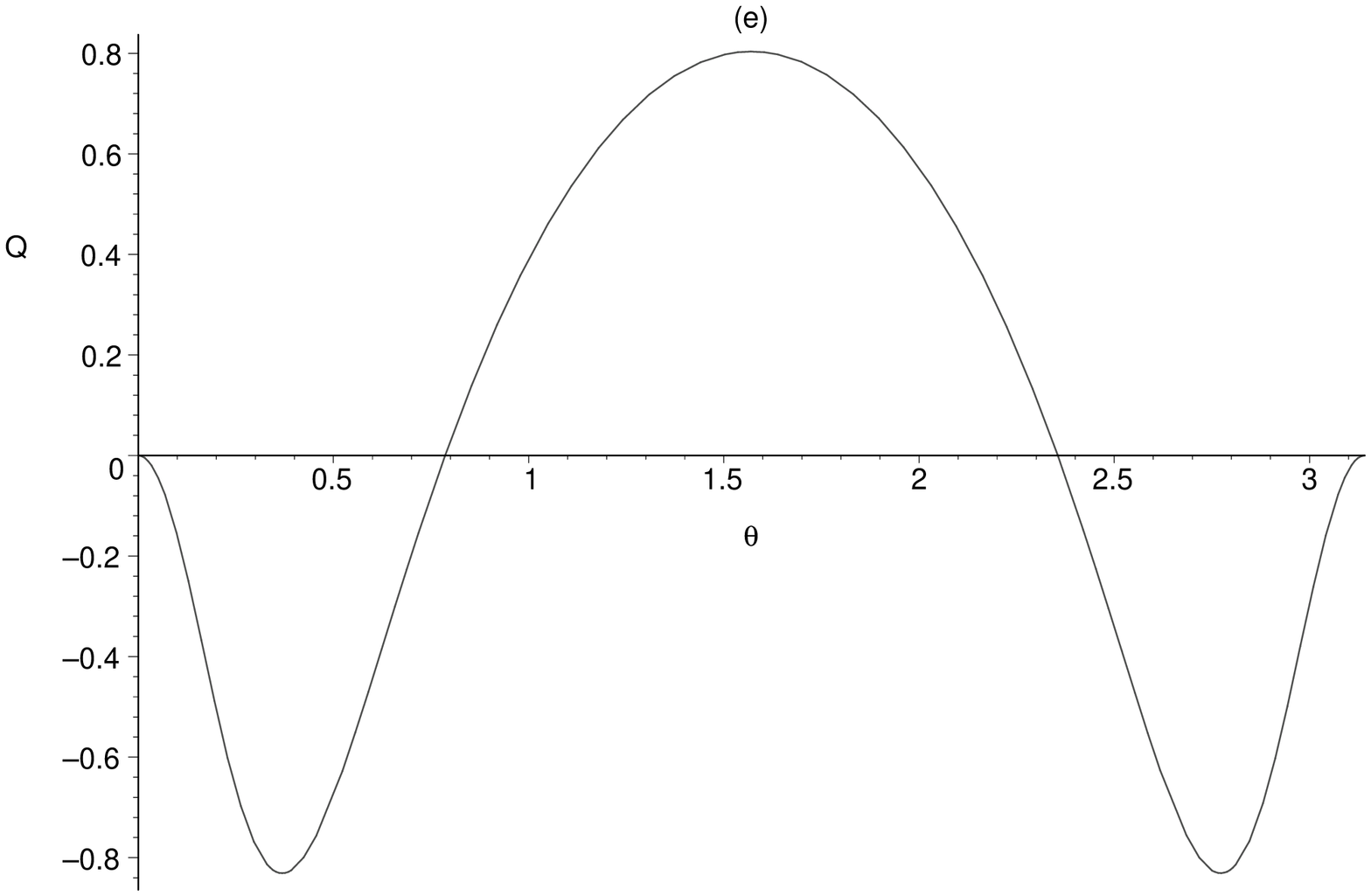}} &
{\includegraphics[width = 1.50in, angle = 270]{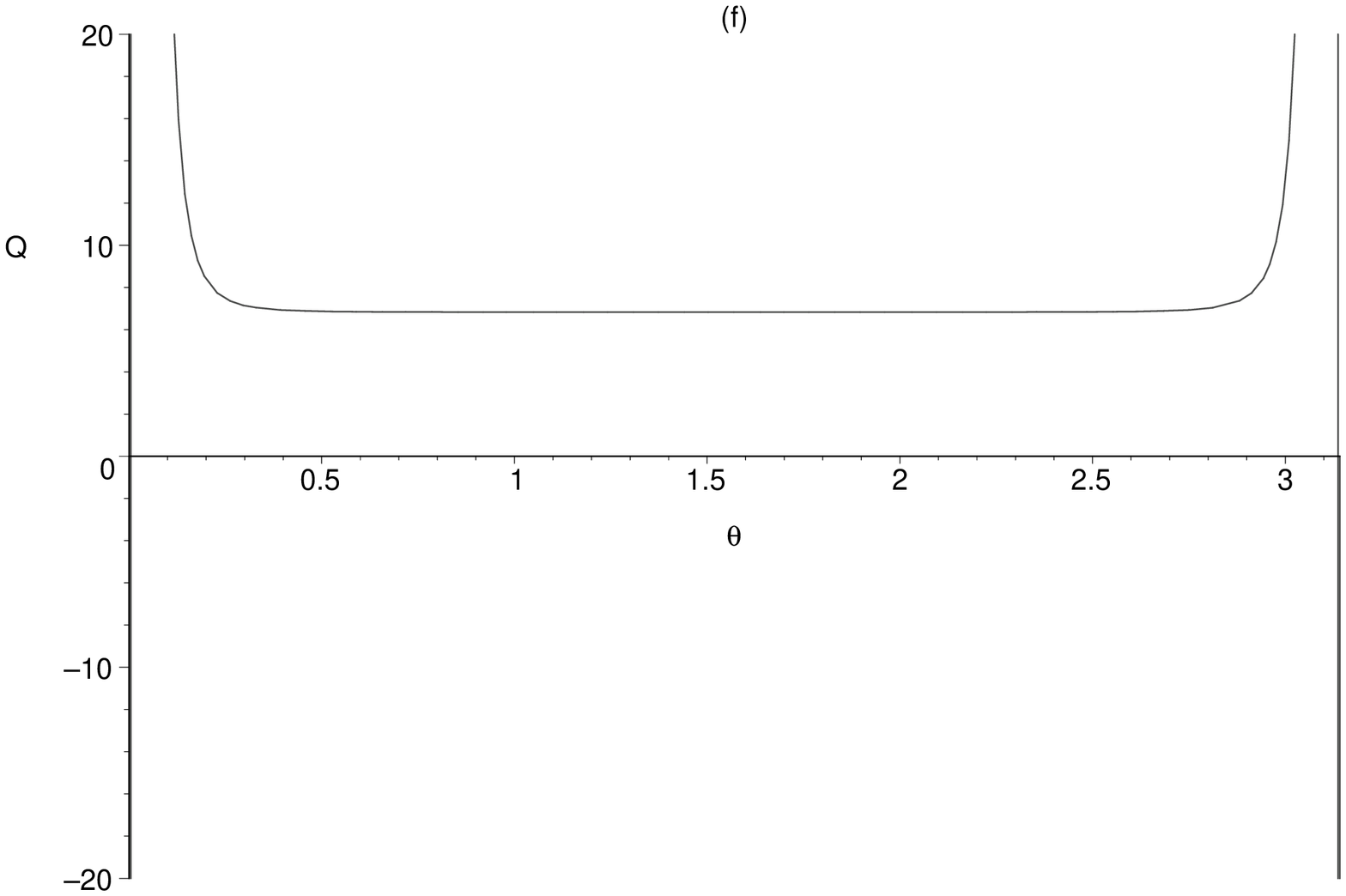}} \\
(e) \  \tau = 0 & (f)\    \tau = +5 \\
{\includegraphics[width = 1.50in, angle = 270]{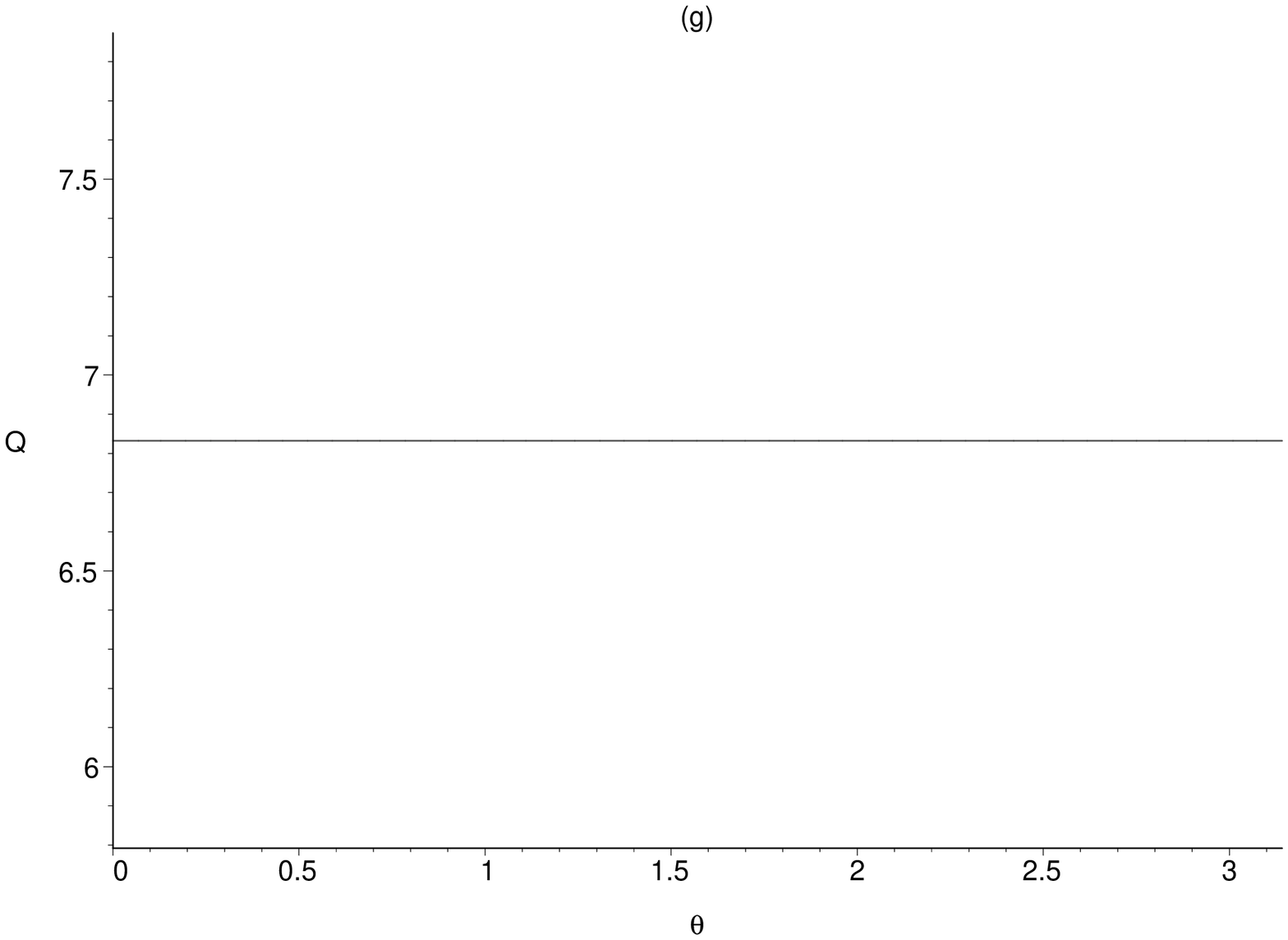}} & 
   \  \\
 (g) \ \tau = +\infty &  \  \cr
\end{array}
\] 
\caption[]{The evolution in $\tau$ of $Q$ 
as a function of $\theta$. }
\end{figure}

\begin{figure}
\begin{center}
\includegraphics[width=.7\textwidth,angle=270]{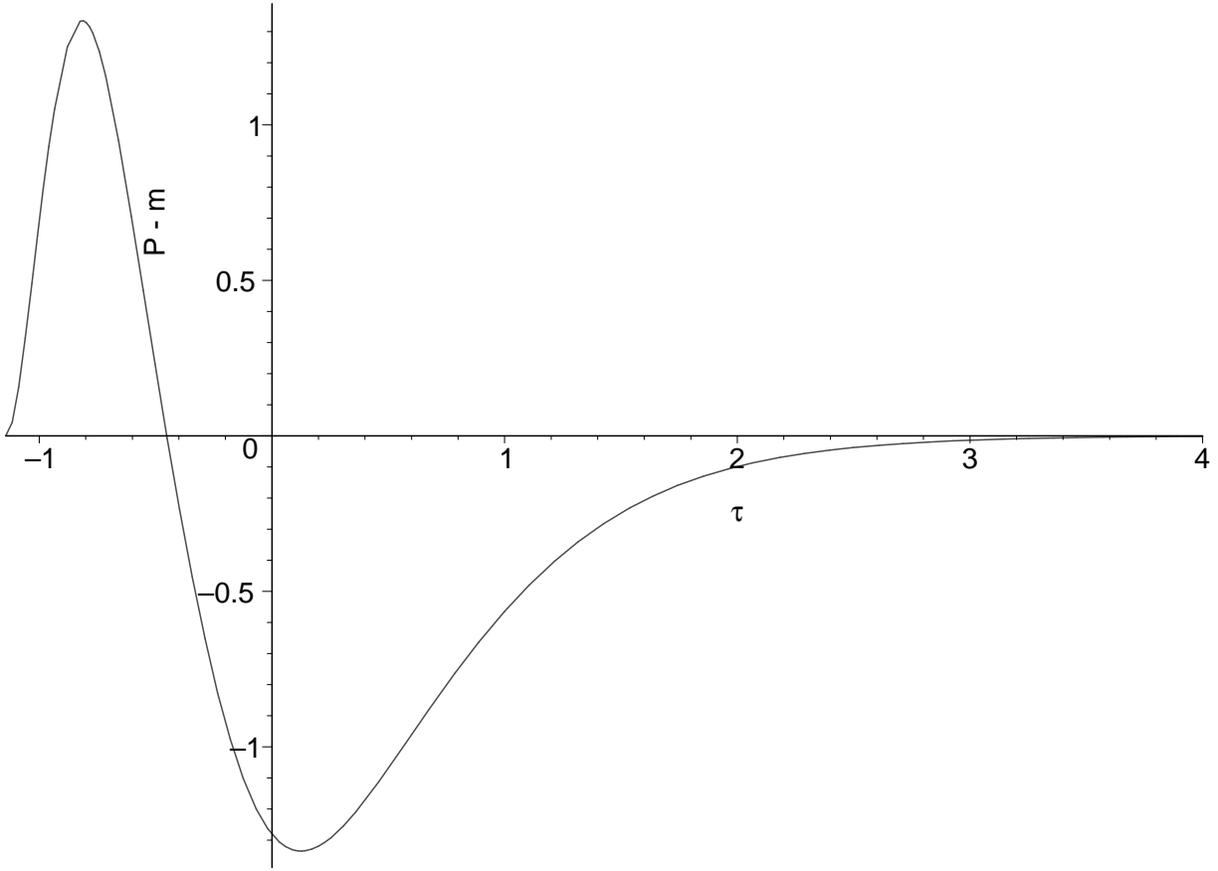}
\end{center}
\caption [] {$P - m$ at $\theta = 0$ versus  $\tau$.}
\end{figure}

\begin{figure}
\begin{center}
\includegraphics[width=.5\textwidth,angle=270]{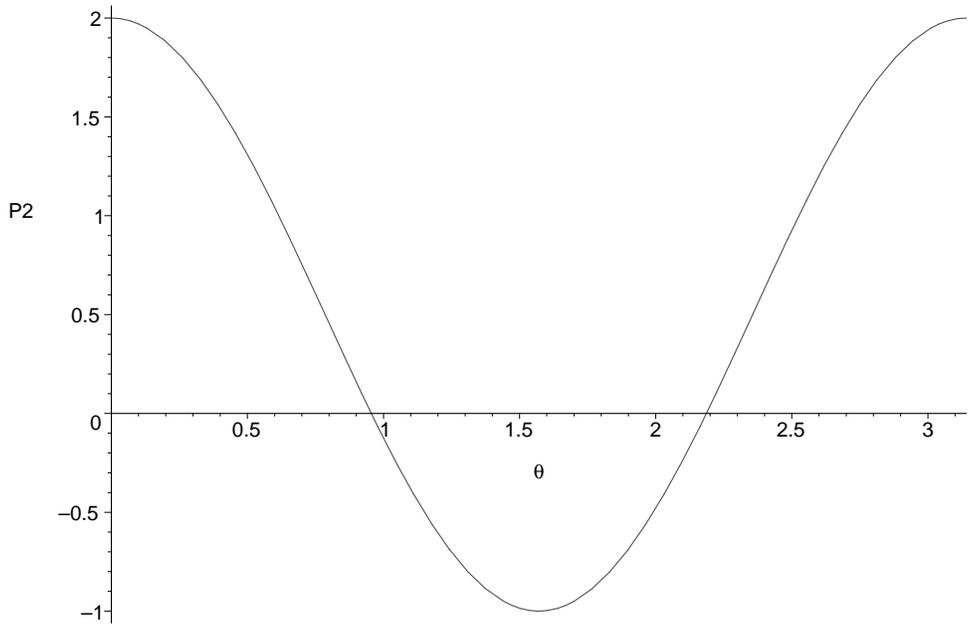}
\end{center}
\caption [] {The Legendre polynomial $P_2 (\cos \theta)$
        versus $\theta$.}
\end{figure}

\begin{figure}
\begin{center}
\includegraphics[width=.7\textwidth,angle=270]{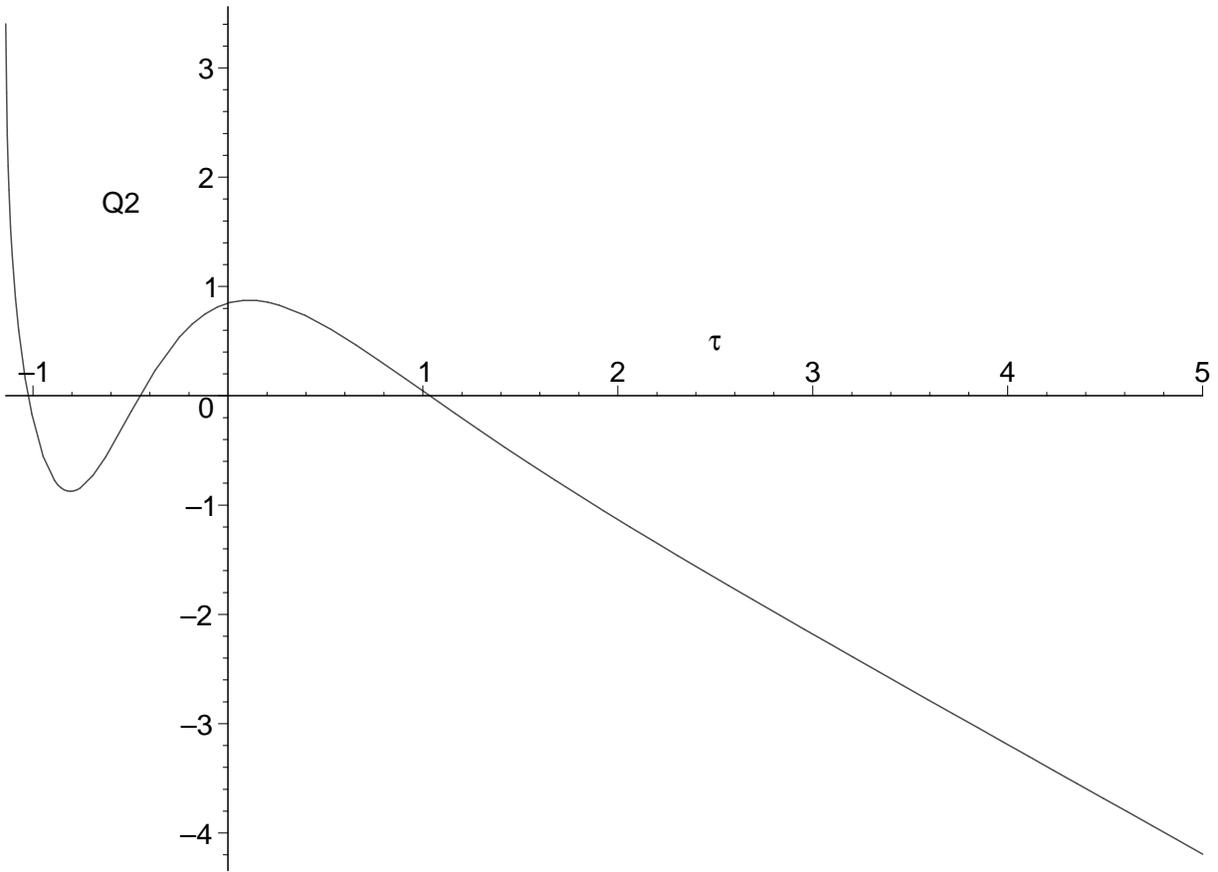}
\end{center}
\caption [] {The Legendre function $Q_2 (\cos e^{-\tau})$
        versus $\tau$.}
\end{figure}

\end{document}